\documentclass[nofootinbib,preprintnumbers,prd,superscriptaddress,twocolumn,showpacs]{revtex4-1}

\usepackage{amsmath}
\usepackage{amsfonts}
\usepackage{amssymb}
\usepackage{graphicx}
\usepackage[titletoc]{appendix}
\usepackage{color}
\usepackage{hyperref}
\usepackage{cleveref}
\usepackage[rightcaption]{sidecap}
\usepackage{subfigure}
\usepackage{comment}
\usepackage{soul}
\usepackage{cancel}
\usepackage{dcolumn}

\usepackage{array}
\usepackage{ctable}
\usepackage{multirow}
\usepackage{siunitx}
\usepackage{longtable}
\usepackage{tabularx}
\usepackage{booktabs}
\usepackage{float}

\graphicspath{{Graphics/}}

\def\be{\begin{equation}}
\def\ee{\end{equation}}
\def\bea{\begin{eqnarray}}
\def\eea{\end{eqnarray}}

\def\nat{Nature}
\def\prl{Phys. Rev. Lett.}

\def\prd{Phys. Rev. D}
\def\mnras{MNRAS}

\def\apj{ApJ}
\def\apjl{ApJ Lett.}
\def\apjs{ApJ Suppl. Ser.}

\def\aap{A\&A}

\def\araa{Annual Rev. of Astron. Astrophys.}

\def\jcap{JCAP}
\def\nar{New Astron. Rev.}

\definecolor{vividviolet}{rgb}{0.62, 0.0, 1.0}
\definecolor{amaranth}{rgb}{0.9, 0.17, 0.31}
\definecolor{palatinateblue}{rgb}{0.15, 0.23, 0.89}
\definecolor{brightpink}{rgb}{1.0, 0.0, 0.5}
\definecolor{cornflowerblue}{rgb}{0.39, 0.58, 0.93}
\definecolor{deepcarminepink}{rgb}{0.94, 0.19, 0.22}
\definecolor{radicalred}{rgb}{1.0, 0.21, 0.37}

\hypersetup{ linktoc=all,
    colorlinks, linkcolor={palatinateblue},
    citecolor={brightpink}, urlcolor={amaranth}
}

\DeclareUnicodeCharacter{2212}{\ensuremath{-}}

\begin{document}

\title{Cosmological constraints from calibrated $E_p-E_{iso}$ gamma-ray burst correlation by using DESI 2024 data release}

\author{Anna Chiara Alfano}
\email{a.alfano@ssmeridionale.it}
\affiliation{Scuola Superiore Meridionale, Largo S. Marcellino 10, 80138 Napoli, Italy.}
\affiliation{Istituto Nazionale di Fisica Nucleare (INFN), Sezione di Napoli Complesso Universitario Monte S. Angelo, Via Cinthia 9 Edificio G, 80138 Napoli, Italy.}

\author{Orlando Luongo}
\email{orlando.luongo@unicam.it}
\affiliation{Universit\`a di Camerino, Divisione di Fisica, Via Madonna delle carceri 9, 62032 Camerino, Italy.}
\affiliation{SUNY Polytechnic Institute, 13502 Utica, New York, USA.}
\affiliation{INFN, Sezione di Perugia, Perugia, 06123, Italy.}
\affiliation{INAF - Osservatorio Astronomico di Brera, Milano, Italy.}
\affiliation{Al-Farabi Kazakh National University, Al-Farabi av. 71, 050040 Almaty, Kazakhstan.}

\author{Marco Muccino}
\email{marco.muccino@lnf.infn.it}
\affiliation{Universit\`a di Camerino, Divisione di Fisica, Via Madonna delle carceri 9, 62032 Camerino, Italy.}
\affiliation{Al-Farabi Kazakh National University, Al-Farabi av. 71, 050040 Almaty, Kazakhstan.}
\affiliation{ICRANet, Piazza della Repubblica 10, 65122 Pescara, Italy.}

\begin{abstract}
Recent outcomes by the DESI Collaboration have shed light on a possible slightly evolving dark energy, challenging the standard $\Lambda$CDM paradigm. To better understand dark energy nature, high-redshift observations like gamma-ray burst data become essential for mapping the universe expansion history, provided they are calibrated with other probes. To this aim, we calibrate the $E_p-E_{iso}$ (or Amati) correlation through model-independent B\'ezier interpolations of the updated Hubble rate and the novel DESI data sets. More precisely, we provide two B\'ezier calibrations: i) handling the entire DESI sample, and ii) excluding the point at $z_{eff}=0.51$, criticized by the recent literature. In both the two options, we let the comoving sound horizon at the drag epoch, $r_d$, vary in the range $r_d \in [138, 156]$ Mpc. The Planck value is also explored for comparison. By means of the so-calibrated gamma-ray bursts, we thus constrain three dark energy frameworks, namely the standard $\Lambda$CDM, the $\omega_0$CDM and the  $\omega_0\omega_1$CDM models, in both spatially flat and non-flat universes. To do so, we worked out Monte Carlo Markov chain analyses, making use of the Metropolis-Hastings algorithm. Further, we adopt model selection criteria to check the statistically preferred cosmological model finding a preference towards the concordance paradigm  only whether the spatial curvature is zero. Conversely, and quite interestingly, the flat $\omega_0$CDM and both the cases,  flat/non-flat, $\omega_0\omega_1$CDM model, leave evidently open the chance that dark energy evolves at higher redshifts.
\end{abstract}



\maketitle
\tableofcontents

\section{Introduction}

Recent cosmological observations from DESI mission seem to indicate that dark energy may slightly evolve \cite{2024arXiv240403002D}. This result appears in tension with previous expectations, where the late-time universe appeared very-well modelled by an anti-gravitating cosmological constant, $\Lambda$ \cite{1992ARA&A..30..499C, 2001LRR.....4....1C, 2003RvMP...75..559P}.

Accordingly, the concordance background  assumes that dark energy exhibits an exotic equation of state, sufficiently negative to push the universe to accelerate \cite{2008ARA&A..46..385F, 2001NewAR..45..235L, 2003PhT....56d..53P}. Consequently, numerous dark energy scenarios have been investigated, during the last decades, to clarify whether dark energy evolves in time or not \cite{2001IJMPD..10..213C, 2003PhRvL..90i1301L, 2006IJMPD..15.1753C}.

Statistically speaking, the $\Lambda$CDM model was always favoured with respect to other approaches, and so the preliminary release from the DESI collaboration suggests a potential new physics reinforcing the well-known cosmological tensions \cite{2022JHEAp..34...49A, 2021APh...13102604D,  2021APh...13102605D} and the thorny theoretical inconsistencies of the $\Lambda$CDM paradigm\footnote{For an alternative to the standard cosmological paradigm that solves the cosmological constant problem, mimicking the cosmological constant through an effective bare term, see e.g. \cite{2018PhRvD..98j3520L}.} \cite{2006IJMPD..15.1753C, 2014EPJC...74.3160V}.

For all these reasons, the need of high-redshift indicators is absolutely crucial, since supernova data have also been criticized \cite{2010RAA....10.1195G, 2020ApJ...902...14S, 2021CQGra..38o4005R, 2019arXiv190500221R} and does not fix the Hubble constant $H_0$ by itself. In other words, adopting high-redshift data sets may be crucial to clarify the nature of dark energy, in order to  shed light on its possible evolution and, so, checking the goodness of DESI outcomes.

In this respect, gamma-ray bursts (GRBs), appear promising, as high-energy astrophysical sources of $\gamma$-rays reaching up to redshifts $z\simeq9$ \cite{2009Natur.461.1258S, 2011ApJ...736....7C}. Even though appealing, GRBs are absolutely far from being standard candles as due to the \emph{circularity problem}, see e.g. \cite{2021Galax...9...77L}, in which calibration requires cosmological information that limits the use of such indicators in cosmological backgrounds. Even though severe criticisms toward the  calibration procedures have been also raised \cite{2008MNRAS.391..411B, 2023JCAP...07..021K, 2008PhRvD..78l3532W}, these objects remain essential to frame the universe out, albeit, quite strangely, seem to indicate larger values of the mass parameter, typically constrained to higher values with respect to other cosmological probes \cite{amati2019addressing, 2021JCAP...09..042K, 2021MNRAS.503.4581L, 2024JHEAp..42..178A}.

Consequently, standardizing GRBs, with the aim of finding new distance indicators, turns out to be a very hard task, also in view of the fact that several GRB correlations have bee proposed \cite{2013IJMPD..2230028A, 2015A&A...582A.115I, 2004ApJ...609..935Y, 2022MNRAS.512..439C, 2004ApJ...616..331G, 2005ApJ...633..611L}. To this end, model-independent approaches that utilize GRB data may be crucial to determine whether dark energy behaves as a cosmological constant or evolves over time \cite{2021ApJ...908..181M, 2023MNRAS.523.4938M, 2007APh....27..113H, 2024JHEAp..42..178A}. For example, the cosmographic approach \cite{1976Natur.260..591H, 2004CQGra..21.2603V, visser2005cosmography, 2016IJGMM..1330002D} provides a method based on Taylor expansions that can feature the universe evolution without postulating the model \emph{a priori}. Although widely-adopted \cite{2024arXiv240407070L,2022MNRAS.509.5399C, 2023PDU....4201298A}, model-independent methods do not provide smaller values of mass densities that still appear quite large.

Motivated by the above theoretical context, we here develop a set of  analyses that makes use of the most recent DESI 2024 data release to calibrate GRBs. To do so, we resort to B\'ezier polynomials, first introduced in Ref. \cite{amati2019addressing}. Our recipe consists in writing both the luminosity distance and Hubble rate appearing in the distances of the DESI sample in terms of B\'ezier parametric curves. This permits to have a fully model-independent method to test a given cosmological model. To work the  B\'ezier interpolation out, we calibrate the $E_p-E_{iso}$ (or Amati) correlation, through the novel DESI data points and the most recent observational Hubble data (OHD). The strategy consists in two main methods, the first handling the entire DESI data set, and the second excluding the point placed at $z_{eff}=0.51$, in fulfilment with recent literature that remarked a pathological behavior of this point \cite{2024arXiv240408633C}.
For both methods, we ensure the range $r_d \in [138, 156]$ Mpc, performing the numerical analyses at fixed steps $\delta r_d=2$ Mpc; the Planck value is also explored for comparison.
Hence, Monte Carlo Markov chain (MCMC) fits, using the Metropolis-Hastings algorithm \cite{1953JChPh..21.1087M, 1970Bimka..57...97H}, are involved to determine the GRB correlation coefficients and, afterwards, the cosmological parameters using, as claimed above, the $E_p-E_{iso}$, or Amati, correlation, relating the emitted isotropic energy,  $E_{iso}$,  with the peak energy, $E_p$ \cite{2002A&A...390...81A, 2013IJMPD..2230028A}.
Our overall analysis is modelled for flat and non-flat $\Lambda$CDM, $w_0$CDM and $w_0w_1$CDM models.
Our findings are finally compared with statistical criteria, to check their goodness and, also, to establish the most favored cosmological puzzle.
The flat $\Lambda$CDM model appears to be favoured, whereas its non-flat extension and the flat $w_0$CDM model cannot be fully excluded, representing valid alternatives to the standard scenario.

The paper is structured as follows. In Sec.~\ref{sec2}, we introduce the model-independent Bez\'ier calibration approach. In Sec.~\ref{sec3} we describe our calibration technique for the $E_p$-$E_{iso}$ correlation, whereas, in Sec.~\ref{sec4}, we show the results of our MCMC computations for the above mentioned three different cosmological models. There, we also compute, for each scenario, the corresponding non-flat extension. Finally, Sec.~\ref{conc} deals with conclusions and perspectives to the present work.

\section{The B\'ezier interpolation}\label{sec2}

B\'ezier polynomials represent a useful tool to get constraints on cosmological parameters through GRBs correlations without assuming a cosmological model \textit{a priori}. Interpolating OHD and baryonic acoustic oscillations (BAO) data sets through B\'ezier parametric curves is widely used in the literature \citep{amati2019addressing,2020A&A...641A.174L, 2021MNRAS.503.4581L, 2023MNRAS.523.4938M, 2023MNRAS.518.2247L, 2023arXiv231105324A, 2024JHEAp..42..178A}. We construct specific B\'ezier polynomials of order $n$ for both OHD and BAO data sets.
\begin{enumerate}
\item [-] \textbf{OHD data.} The B\'ezier curve for this data set is
\begin{equation}\label{bezierOHD}
        H_n(x)=\sum^{n}_{i=0} g_\alpha\alpha_i\left[n!\frac{x^i}{i!}\frac{(1-x)^{n-i}}{(n-i)!}\right],
\end{equation}
where $g_\alpha= 100\
\text{km}\cdot\text{s}^{-1}\cdot\text{Mpc}^{-1}$ is a re-scaling factor, $\alpha_i$ are the so-called \textit{B\'ezier coefficients} where the coefficient $\alpha_0$ is identified with the reduced Hubble constant $h_0\equiv H_0/g_\alpha$, $0\leq x\equiv z/z^{max}_O \leq 1$, and $z^{max}_O$ is the maximum redshift of the OHD catalog. At order $n=2$, Eq.~\eqref{bezierOHD} behaves as a non-linear monotonic growing function in agreement with the behaviour of the Hubble rate. This permits us to approximate $H(z)$ with $H_2(x)$ \citep{2021MNRAS.503.4581L}.
\item [-] \textbf{BAO data.} The B\'ezier curve in this case is
\begin{equation}\label{bezierBAO}       D^2_m(y)=\sum^{m}_{j=0}g_\beta\beta_j\left[m!\frac{y!}{j!}\frac{(1-y)^{m-j}}{(m-j)!}\right],
\end{equation}
where $g_\beta=10^6\ \text{Mpc}^2$ is a re-scaling factor, $\beta_j$ are the \textit{B\'ezier coefficients} for BAO data and $0 \leq y\equiv z/z_B^{max} \leq 1$ where $z^{max}_B$ is the maximum redshift in the BAO catalog. At order $m=3$, Eq.~\eqref{bezierBAO} behaves as non-linear monotonic growing function and the constant term at $j=0$ is identically set equal to zero since, by definition of distance, $D^2_m(0)\equiv0$. In this way we can approximate the luminosity distance $D_L^2(z)$ with $D^2_{13}(y)$, where the subscripts indicate that the interpolation spans the terms $j=\{1, 2, 3\}$ \citep{2023MNRAS.518.2247L}.
\end{enumerate}

To derive the B\'ezier coefficients, we perform MCMC simulations through a Python code, adopting the Metropolis-Hastings algorithm \citep{1953JChPh..21.1087M, 1970Bimka..57...97H}, and maximize the total log-likelihood
\begin{equation}\label{loglikelihoods}
\ln\mathcal{L}_1=\ln\mathcal{L}_{O}+\ln\mathcal{L}_{D},
\end{equation}
where $\ln\mathcal{L}_O$ is the log-likelihood associated with the OHD data set, while $\ln\mathcal{L}_{D}$ is the log-likelihood associated with the DESI-BAO data set. elow, we describe each contribution to the total log-likelihood in Eq.~\eqref{loglikelihoods}

\begin{enumerate}
    \item [-] \textbf{OHD data.} We consider the updated OHD catalog of $N_O = 34$ data points spanning in a redshift interval $z\in [0.0708,1.965]$. This catalog is derived using the so-called \textit{cosmic chronometers} approach which gives model-independent results for the Hubble rate $H(z)$. The idea behind this approach is to consider the relation between the scale factor $a=(1+z)^{-1}$ and the relation between the spectropic observations of the difference $\Delta z$ between galaxies formed at the same time $\Delta t$ who evolve passively acting as \textit{cosmic chronometers} \citep{2002ApJ...573...37J, 2018ApJ...868...84M}.

    \begin{table}
\centering
\setlength{\tabcolsep}{2.4em}
\renewcommand{\arraystretch}{1.1}
   \begin{tabular}{lcc}
   \hline\hline
    $z$     &$H(z)$ &  Refs. \\
            &(km/s/Mpc)&\\
    \hline
    0.07  & $69.0  \pm 19.6\pm 12.4$ & \cite{Zhang2014} \\
    0.09    & $69.0  \pm 12.0\pm 11.4$  & \cite{Jimenez2002} \\
    0.12    & $68.6  \pm 26.2\pm 11.4$  & \cite{Zhang2014} \\
    0.17    & $83.0  \pm 8.0\pm 13.1$   & \cite{Simon2005} \\
    0.1791   & $75.0  \pm 3.8\pm 0.5$   & \cite{Moresco2012} \\
    0.1993   & $75.0  \pm 4.9\pm 0.6$   & \cite{Moresco2012} \\
    0.20    & $72.9  \pm 29.6\pm 11.5$  & \cite{Zhang2014} \\
    0.27    & $77.0  \pm 14.0\pm 12.1$  & \cite{Simon2005} \\
    0.28    & $88.8  \pm 36.6\pm 13.2$  & \cite{Zhang2014} \\
    0.3519   & $83.0  \pm 13.0\pm 4.8$  & \cite{Moresco2016} \\
    0.3802  & $83.0  \pm 4.3\pm 12.9$  & \cite{Moresco2016} \\
    0.4     & $95.0  \pm 17.0\pm 12.7$  & \cite{Simon2005} \\
    0.4004  & $77.0  \pm 2.1\pm 10.0$  & \cite{Moresco2016} \\
    0.4247  & $87.1  \pm 2.4\pm 11.0$  & \cite{Moresco2016} \\
    0.4497  & $92.8  \pm 4.5\pm 12.1$  & \cite{Moresco2016} \\
    0.47    & $89.0\pm 23.0\pm 44.0$     & \cite{2017MNRAS.467.3239R}\\
    0.4783  & $80.9  \pm 2.1\pm 8.8$   & \cite{Moresco2016} \\
    0.48    & $97.0  \pm 62.0\pm 12.7$  & \cite{Stern2010} \\
    0.5929   & $104.0 \pm 11.6\pm 4.5$  & \cite{Moresco2012} \\
    0.6797    & $92.0  \pm 6.4\pm 4.3$   & \cite{Moresco2012} \\
    0.75    & $98.8\pm33.6$     & \cite{2022ApJ...928L...4B}\\
    0.7812   & $105.0 \pm 9.4\pm 6.1$  & \cite{Moresco2012} \\
    0.80    & $113.1\pm15.1\pm 20.2$    & \cite{2023ApJS..265...48J}\\
    0.8754   & $125.0 \pm 15.3\pm 6.0$  & \cite{Moresco2012} \\
    0.88    & $90.0  \pm 40.0\pm 10.1$  & \cite{Stern2010} \\
    0.9     & $117.0 \pm 23.0\pm 13.1$  & \cite{Simon2005} \\
    1.037   & $154.0 \pm 13.6\pm 14.9$  & \cite{Moresco2012} \\
     1.26    & $135.0\pm 65.0$          & \cite{Tomasetti2023}\\
    1.3     & $168.0 \pm 17.0\pm 14.0$  & \cite{Simon2005} \\
    1.363   & $160.0 \pm 33.6$  & \cite{Moresco2015} \\
    1.43    & $177.0 \pm 18.0\pm 14.8$  & \cite{Simon2005} \\
    1.53    & $140.0 \pm 14.0\pm 11.7$  & \cite{Simon2005} \\
    1.75    & $202.0 \pm 40.0\pm 16.9$  & \cite{Simon2005} \\
    1.965   & $186.5 \pm 50.4$  & \cite{Moresco2015} \\
\hline
\end{tabular}
\caption{The updated OHD catalog. The columns list the redshifts, the values of $H(z)$ with statistical and systematic errors, and references, respectively.}
\label{tab1}
\end{table}

The 34 OHD data points are listed inside Table~\ref{tab1} and their log-likelihood takes the form
\begin{equation}
     \ln\mathcal{L}_O = -\frac{1}{2}\sum^{N_O}_{i=1}\left\{\left[\frac{H_i-H_2(z_i)}{\sigma^2_{H_i}}\right]^2+\ln\left(2\pi\sigma^2_{H_i}\right)\right\},
\end{equation}
where $H_i$ are the observational data points with errors $\sigma_{H_i}=\sqrt{\left[\sigma^{stat}_{H_i}\right]^2+\left[\sigma^{sys}_{H_i}\right]^2}$  while $H_2(z_i)$ is the reconstructed Hubble rate up to the second order.
\begin{table}
\centering
\setlength{\tabcolsep}{.3em}
\renewcommand{\arraystretch}{1.1}
\begin{tabular}{lcccc}
\hline
Tracer     & $z_{eff}$ & $D_M/r_d$ & $D_H/r_d$ & $D_V/r_d$ \\
\hline
BGS & $0.30$ & $-$ & $-$ & $7.93\pm 0.15$  \\
LRG1 & $0.51$ & $13.62\pm 0.25$ & $20.98\pm 0.61$ & $-$  \\
LRG2 & $0.71$ & $16.85\pm 0.32$ & $20.08\pm 0.60$ & $-$  \\
LRG3+ELG1 & $0.93$ & $21.71\pm 0.28$ & $17.88\pm 0.35$ & $-$ \\
ELG2 & $1.32$ & $27.79\pm 0.69$ & $13.82\pm 0.42$ & $-$  \\
QSO & $1.49$ & $-$ & $-$ & $26.07\pm 0.67$  \\
Lya QSO & $2.33$ & $39.71\pm 0.94$ & $8.52\pm 0.17$ & $-$ \\
\hline
\end{tabular}
\caption{Values with associated errors of the distances for the six tracers at the effective redshift $z_{eff}$ \cite{2024arXiv240403002D}.}
\label{tab:DESIBAO}
\end{table}
\item [-] \textbf{DESI-BAO data.} The newly-released BAO data from the DESI collaboration consist of a sample of six tracers: bright galaxy survey (BGS), luminous red galaxies (LRG), emission line galaxies (ELG), quasars (QSO), Lyman-$\alpha$ forest quasars (Lya QSO) and a combination of LRG+ELG. These tracers span a redshift interval $z\in[0.1, 4.2]$ and through them it is possible to derive the values of the transverse comoving distance $D_M(z)/r_d$, the Hubble rate distance $D_H(z)/r_d$ and the angle-average distance $D_V(z)/r_d$, defined as
\begin{subequations}\label{desidist}
    \begin{align}
        &\frac{D_M(z)}{r_d}=\frac{D_L(z)}{r_d}(1+z)^{-1},&\\
        &\frac{D_H(z)}{r_d} = \frac{c}{r_d H(z)},&\\
        &\frac{D_V(z)}{r_d} = \frac{1}{r_d}\left[\frac{cz D^2_L(z)}{(1+z)^2H(z)}\right]^{1/3},&
    \end{align}
\end{subequations}
where $r_d$ is the comoving sound horizon at the drag epoch which we assume to span in the interval $r_d\in[138,156]$ Mpc in which the expectations for both the DESI and Planck satellite fall. The values of the distances defined in Eqs.~\eqref{desidist} divided in seven redshift bins \cite{2024arXiv240403002D}
for the six tracers are listed in Table~\ref{tab:DESIBAO} and the log-likelihood takes the form
\begin{equation}
    \ln\mathcal{L}_{D}=-\frac{1}{2}\sum^{N_{DB}}_{i=1}\left\{\left[\frac{D_i-D(z_i)}{\sigma^2_{D_i}}\right]+\ln\left(2\pi \sigma^2_{D_i}\right)\right\},
\end{equation}
where we identify $D_i=\{D_M/r_d, D_H/r_d, D_V/r_d\}$ while $D(z_i)$ are the distance ratios written in terms of B\'ezier curves.
\end{enumerate}

We computed the values of B\'ezier coefficients in two different ways, listed below.
\begin{enumerate}
    \item We considered the complete data set from DESI and let $r_d$ vary in the range $r_d\in[138, 156]$ Mpc at steps of $2$ Mpc; further, for comparison, we also performed a fit with the value $r_d=147.09\pm 0.26$ from the \citet{2020A&A...641A...6P}.
    \item We performed the same scan in $r_d$ as before, but excluding the  LRG1 data point at $z_{eff}=0.51$, in line with Ref. \cite{2024arXiv240408633C} where the authors show a higher value of $\Omega_m$ for this data point.
\end{enumerate}

Finally, after computing our simulations the values of B\'ezier coefficients $\alpha_i$ and $\beta_j$ for each point discussed above are listed in Tables~\ref{tab:DESIBAOrdvar}--\ref{tab:DESIBAOrdvarwithout051} in Appendix \ref{appendix1} while Fig.~\ref{fig:1} shows the best-fitting B\'ezier curves compared with the flat $\Lambda$CDM scenario \cite{2020A&A...641A...6P}.

\begin{figure*}
\centering
{\hfill
\includegraphics[width=0.49\hsize,clip]{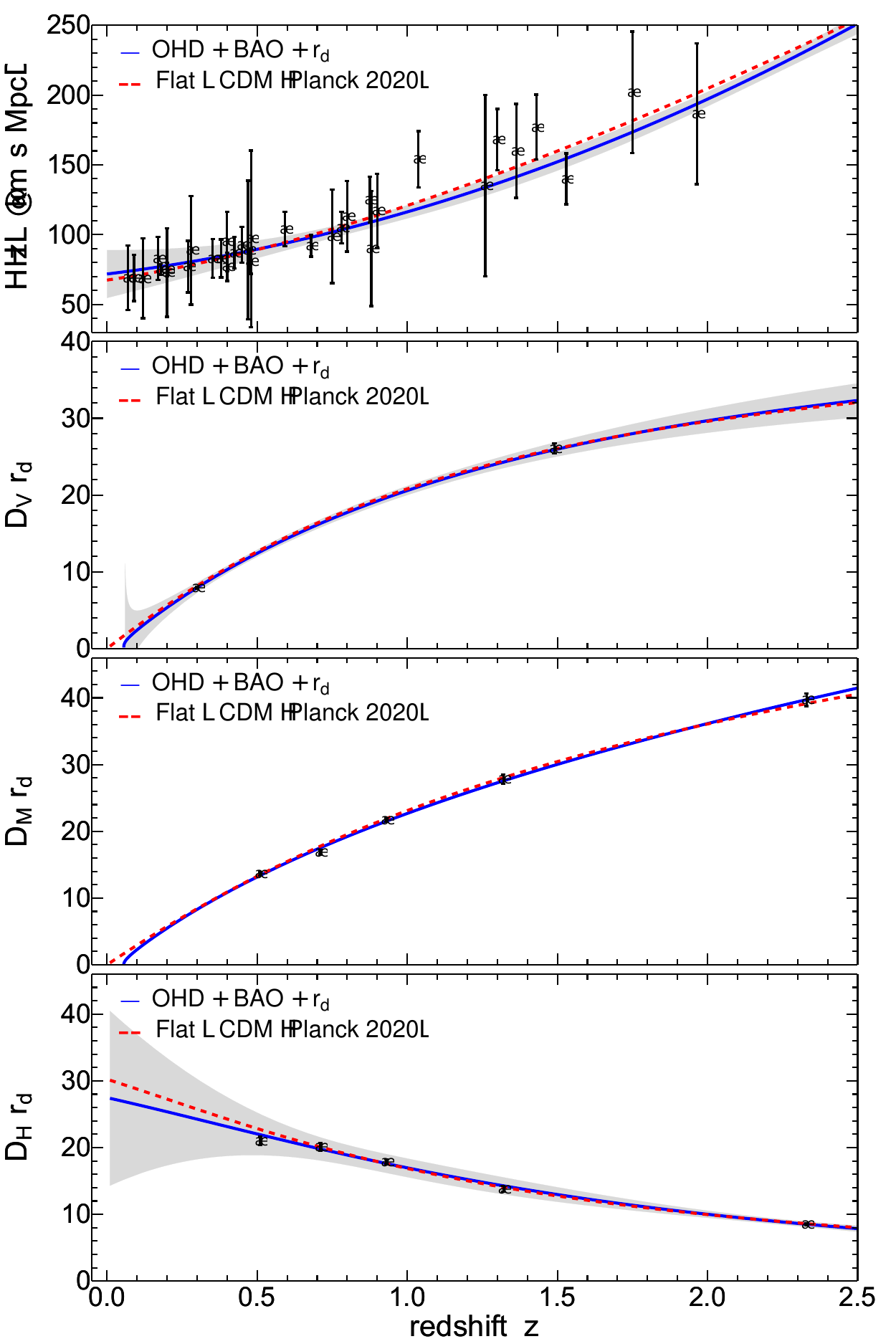}
\hfill
\includegraphics[width=0.49\hsize,clip]{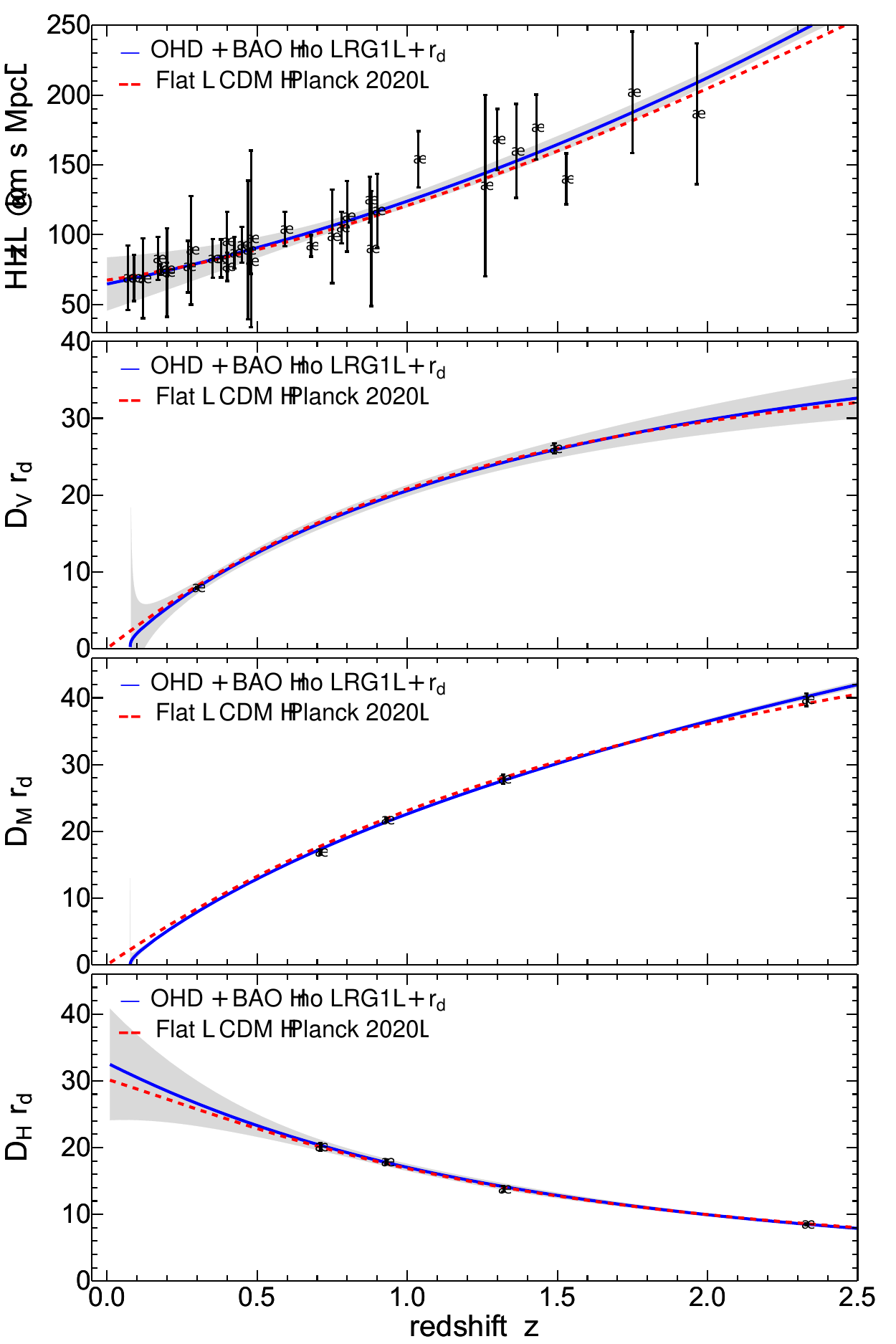}
\hfill}
\caption{Bézier parametric curves of the Hubble rate $H(z)$, the angle-average distance $D_V(z)/r_d$, the transverse comoving distance $D_M(z)/r_d$, and the Hubble rate distance $D_H(z)/r_d$ (solid blue line), compared with the flat $\Lambda$CDM model (dashed red line) for the DESI catalog with (left) and without (right) the LRG1 data point. Grey areas refer to the $1$-$\sigma$ error bands.
}
\label{fig:1}
\end{figure*}

\section{Calibrating the $E_p$-$E_{iso}$ correlation via the DESI sample}\label{sec3}

The $E_p$-$E_{iso}$, or Amati, correlation links the isotropic emitted energy $E_{iso}$ with the rest-frame peak energy $E_p=E_p^o(1+z)$, with $E_p^o$ being the obseverd one \cite{2013IJMPD..2230028A},
\begin{equation}
    \log\left(\frac{E_p}{\text{keV}}\right)=b-a\left[52-\log\left(\frac{E_{iso}}{\text{erg}}\right)\right],
\end{equation}
where $E_{iso}$ depends on the assumed cosmological model, leading to a \emph{circularity} problem, through
\begin{equation}\label{eiso}
    E_{iso}=4\pi D^2_L(z, p)S_b (1+z)^{-1},
\end{equation}
where $S_b$ is the bolometric fluence and $D_L$ is the luminosity distance which depends on the redshift $z$ and on the parameters $p$ of the fiducial cosmological model.

We can evade the model-dependency by calibrating $E_{iso}$ through the B\'ezier-reconstructed luminosity distance in Eq.~\eqref{bezierBAO}. Substituting it inside Eq.~\eqref{eiso} gives
\begin{equation}
    E^{cal}_{iso}=4\pi D^2_{13}(z)S_b (1+z)^{-1}.
\end{equation}

We are now able to determine the intercept $b$, the slope $a$ of the $E_p$-$E_{iso}$ correlation and the cosmological parameters of our interest. We consider a data set composed of $118$ GRBs within a redshift range $z\in [0.3399, 8.2]$ \cite{2021JCAP...09..042K}. Following the same recipe for the determination of B\'ezier coefficients we implement a MCMC analysis by maximazing the following log-likelihood
\begin{equation}
    \ln\mathcal{L}_A=\ln\mathcal{L}^{cal}_A+\ln\mathcal{L}^{cos}_A,
\end{equation}
where $\ln\mathcal{L}^{cal}_A$ is the calibration log-likelihood used to estimate the correlation parameters. This is done by considering a sub-sample in the GRB data set at $z\leq z_B^{max}$
\begin{equation}
\ln\mathcal{L}_A^{cal}=-\frac{1}{2}\sum^{N_{cal}}_{j=1}\left\{\left[\frac{Y_j-Y(z_j)}{\sigma_{Y_j}}\right]^2+\ln(2\pi\sigma^2_{Y_j})\right\},
\end{equation}
where $N_{cal}= 65$ are the data points of the calibrated sample, while we introduced the following definitions
\begin{subequations}
    \begin{align}
        &Y_j-Y(z_j)=\log E_p-b+a\left[52-\log\left(\frac{E_{iso}}{\text{erg}}\right)\right],\\
        &\sigma^2_{Y_j}=\sigma^2_{\log E_p}+a^2\sigma^2_{\log E_{iso}}+\sigma^2_{ex},
    \end{align}
\end{subequations}
in which $\sigma^2_{ex}$ is an extrascatter term \cite{2005physics..11182D}.

The log-likelihood that takes into accout the cosmological model is
\begin{equation}
    \ln\mathcal{L}_A^{cos}=-\frac{1}{2}\sum^{N_{cos}}_{j=1}\left\{\left[\frac{\mu_j-\mu(z_j)}{\sigma_{\mu_j}}\right]^2+\ln(2\pi\sigma^2_{\mu_j})\right\},
\end{equation}
where $N_{cos} = 118$ is the number of data points for the total GRBs sample with distance moduli and attached errors, respectively
\begin{subequations}
    \begin{align}
        \mu_j=\mu_0+\frac{5}{2}\left[\frac{\log E_p-b}{a}-\frac{\log S_b}{(1+z)}\right],\\
        \sigma^2_{\mu_j} = \left(\frac{5}{2}\right)^2\left[\frac{1}{a^2}\sigma^2_{\log_{E_p}}+\sigma^2_{\log S_b}+\sigma_{ex}^2\right],
    \end{align}
\end{subequations}
where $\mu_0=25+2.5\left[52-\log\left(4\pi \xi^2\right)\right]$ and $\xi$ converting Mpc to cm while $\mu(z)$ is the theoretical distance moduli
\begin{equation}
    \mu(z)=25+5\log\left[\frac{D_L(z)}{\text{Mpc}}\right],
\end{equation}
with the luminosity distance $D_L$ taking the form
\begin{equation}\label{dlum}
    D_L(z) = \frac{c(1+z)}{100h_0\sqrt{|\Omega_k|}}S_k\left[\int^z_0 \sqrt{|\Omega_k|}\frac{100h_0}{H(z^\prime)} dz^\prime\right],
\end{equation}
where $H(z)$ is the Hubble rate of the cosmological model we want to test and \cite{1995ApJ...450...14G}
\begin{equation}
    S_k(x) = \begin{cases}
    {\sinh(x)}, & \text{for} \quad \Omega_k > 0, \\
        x, & \text{for} \quad \Omega_k = 0,\\
    {\sin(x)}, & \text{for} \quad \Omega_k < 0.
    \end{cases}
\end{equation}

\section{Checking cosmological models with calibrated GRBs }\label{sec4}

We can now use the calibrated $E_p-E_{iso}$ correlation to test cosmological models, directly considering the Hubble rate, $H(z)$, for each dark energy model under exam,
\begin{equation}\label{hrate}
    H(z)=H_0\sqrt{\Omega_m(1+z)^3+\Omega_k(1+z)^2+\Omega_{de}G(z)},
\end{equation}
where $\Omega_{de} = 1-\Omega_m-\Omega_k$ and $G(z)$ depends upon the assumed cosmological model, however being constrained at our time, by $G(0)=1$.

We substitute Eq.~\eqref{hrate} into Eq.~\eqref{dlum} and perform MCMC simulations.
We fix $h_0$ and $r_d$ to the best-fit values of the B\'ezier interpolations in Tables~\ref{tab:DESIBAOrdvar}--\ref{tab:DESIBAOrdvarwithout051}
\begin{subequations}
    \begin{align}
    \label{h0152}
    {\rm With\ LRG1:}\ h_0 = 0.718^{+0.022(0.048)}_{-0.020(0.046)}, \ r_d = 152\,,\\
    \label{h0142}
    {\rm No\ LRG1:}\ h_0 = 0.646^{+0.023(0.054)}_{-0.036(0.068)}, \ r_d = 142\,.
    \end{align}
\end{subequations}

The results of our computations are reported in Tables~\ref{tab:DESIBAOLCDM}-\ref{tab:DESIBAOCPL} where the following priors have been used
\begin{align*}
    &a\in [0., 1.], \quad b\in [1., 3.], \quad \sigma\in [0., 1.],&\\
    &\Omega_k\in [-3., 3.], \quad \Omega_m\in[0., 1.],&\\
    &\omega_0\in [-8., 4.], \quad \omega_1\in [-10., 6.],&
\end{align*}
while Figs.~\ref{fig:2}-\ref{fig:6} in Appendix \ref{appendix1}  portrait the contour plots obtained by modifying the PYGTC code of \citet{Bocquet2016} for both the flat and non-flat scenario.

Before proceeding with the cosmological analyses, it is worth to confront our values of $h_0$, listed in Eqs.~\eqref{h0142}-\eqref{h0152}, with the ones from the \citet{2020A&A...641A...6P}, say $h_0^P$, and from \citet{2022ApJ...934L...7R}, say $h^R_0$.
\begin{itemize}
    \item[-] The whole DESI data set provides values of $h_0$ consistent at $1$-$\sigma$ with $h^R_0 = 0.730\pm 0.010$, while the consistency is only at $2$-$\sigma$ with $h_0^P=0.674\pm 0.005$ from the flat case.
    \item[-] On the other hand, by excluding the LRG1 data point, our inferred $h_0$ agrees at $1$-$\sigma$ with the \citet{2020A&A...641A...6P}, in both flat ($h_0^P=0.674\pm 0.005$) and non-flat ($h_0^P=0.636^{+0.021}_{-0.023}$) scenarios, whereas it is not consistent with Riess' value $h^R_0$.
\end{itemize}

\begin{table*}[t]
\centering
\setlength{\tabcolsep}{1.8em}
\renewcommand{\arraystretch}{1.7}
\begin{tabular}{ccccc}
\hline
$a$ & $b$ & $\sigma$ & $\Omega_k$ & $\Omega_m$ \\
\hline
\multicolumn{5}{c}{With LRG1}\\
\cline{1-5}
$0.713^{+0.049(0.087)}_{-0.057(0.081)}$ & $1.809^{+0.072(0.123)}_{-0.098(0.158)}$ & $0.292^{+0.030(0.052)}_{-0.029(0.046)}$ & $-$ & $0.255^{+0.135(0.241)}_{-0.115(0.161)}$ \\
$0.703^{+0.048(0.092)}_{-0.055(0.081)}$ & $1.831^{+0.080(0.123)}_{-0.091(0.157)}$ & $0.290^{+0.029(0.052)}_{-0.030(0.046)}$ & $0.323^{+0.890(1.755)}_{-0.466(0.682)}$ & $0.197^{+0.140(0.230)}_{-0.149(0.196)}$ \\
\hline
\multicolumn{5}{c}{No LRG1} \\
\cline{1-5}
$0.716^{+0.050(0.085)}_{-0.059(0.090)}$ & $1.837^{+0.076(0.125)}_{-0.091(0.148)}$ & $0.291^{+0.030(0.053)}_{-0.029(0.044)}$ & $-$ & $0.391^{+0.191(0.339)}_{-0.147(0.214)}$ \\
$0.704^{+0.058(0.089)}_{-0.056(0.085)}$ & $1.864^{+0.080(0.131)}_{-0.094(0.146)}$ & $0.289^{+0.032(0.054)}_{-0.031(0.043)}$ & $0.453^{+1.453(2.185)}_{-0.687(0.904)}$ & $0.251^{+0.183(0.299)}_{-0.181(0.251)}$ \\
\hline
\end{tabular}
\caption{Best-fit parameters at $1$-$\sigma$ ($2$-$\sigma$) from the MCMC simulation of $E_p-E_{iso}$ correlation, calibrated with and without the DESI LRG1 data point, for the $\Lambda$CDM model with $\Omega_k= 0$ and $\Omega_k\neq 0$. }
\label{tab:DESIBAOLCDM}
\end{table*}

\subsection{The $\Lambda$CDM model}

Here we consider $G(z)\equiv 1$ inside Eq.~\eqref{hrate}. The corresponding results for this model are shown in Table~\ref{tab:DESIBAOLCDM}, where the best-fit coefficients of the $E_p-E_{iso}$ correlation are consistent with results found in the literature in both cases of a flat and non-flat scenario, \cite{2024JHEAp..42..178A, 2023MNRAS.518.2247L}.

Focusing on the cosmological parameters, we compare the concordance paradigm and its non-flat extension with the values from \citet{2020A&A...641A...6P}, labeled by the superscript $P$, and from the Sloan Digital Sky Survey (SDSS) \cite{2021PhRvD.103h3533A}, denoted by the superscript $S$.

We start with the matter parameter. In the flat case:
\begin{itemize}
    \item[-] considering the whole DESI sample, we inferred a low $\Omega_m$ with large errors that is consistent at $1$-$\sigma$ with $\Omega_m^P = 0.315\pm 0.007$ \cite{2020A&A...641A...6P}, and at $1$-$\sigma$ with the one from the SDSS $\Omega_m^S = 0.299 \pm 0.016$ \cite{2021PhRvD.103h3533A};
    \item[-] excluding LRG1, $\Omega_m$ increases but still agrees at $1$-$\sigma$ with both the above flat-case $\Omega_m^S$ and $\Omega_m^P$.
\end{itemize}

In the non-flat scenario:
 \begin{itemize}
     \item[-] considering the whole DESI sample, we found a low $\Omega_m$ that agrees only at $2$-$\sigma$ with $\Omega_m^P=0.348^{+0.013}_{-0.014}$ \cite{2020A&A...641A...6P} and at $1$-$\sigma$ with $\Omega_m^S=0.285^{+0.168}_{-0.170}$ \cite{2021PhRvD.103h3533A}\footnote{The value of $\Omega_m^P$ was inferred by considering $\Omega_bh^2 = 0.02249\pm 0.00016$ and $\Omega_ch^2 = 0.1185\pm 0.0015$ taken from the \citet{2020A&A...641A...6P}. The value of $\Omega_m^S$ was inferred by considering $\Omega_{de} = 0.636^{+0.085}_{-0.070}$ and $\Omega_k = 0.079^{+0.083}_{-0.100}$ from the SDSS \cite{2021PhRvD.103h3533A}.}.
     \item[-] excluding LRG1, also in this case $\Omega_m$ increases and is in agreement at $1$-$\sigma$ with both the above non-flat determinations  $\Omega_m^P$ and $\Omega_m^S$.
 \end{itemize}

It is worth noticing that the higher values of $\Omega_m$ inferred from the DESI catalog without LRG1 are due to the lower value of $h_0$ extracted by the calibration with this smaller data set. This is a direct consequence of the well-known $h_0$--$\Omega_m$ degeneracy \cite{1997ApJ...488....1Z, 2001ApJ...549..669H, 2002MNRAS.337.1068P}. Moreover, it is also well-established that GRBs provide mass estimates which are higher than those extracted by other probes \cite{sym12071118, amati2019addressing}. However, this trend is not confirmed when considering the whole DESI data set, hinting that the LRG1 point may cancel out the effect of the GRB data.

Finally, we focus our attention on the value of curvature parameter $\Omega_k$ inferred from our MCMC simulations:
 \begin{itemize}
     \item[-] considering the whole DESI sample, our $\Omega_k$ is consistent at $1$-$\sigma$ with both the values from the SDSS
     $\Omega_k^S=0.079^{+0.083}_{-0.100}$ and Planck $\Omega_k^P=-0.011^{+0.013}_{-0.012}$;
     \item[-] excluding LRG1, our value of $\Omega_k$ is in agreement at $1$-$\sigma$ with $\Omega_k^S$ and $\Omega_k^P$.
 \end{itemize}

 \subsection{The $\omega_0$CDM model}

\begin{table*}[t]
\centering
\setlength{\tabcolsep}{0.8em}
\renewcommand{\arraystretch}{1.7}
\begin{tabular}{cccccc}
\hline
$a$ & $b$ & $\sigma$ & $\Omega_k$ & $\Omega_m$ & $\omega_0$\\
\hline
\multicolumn{6}{c}{With LRG1} \\
\cline{1-6}
$0.715^{+0.055(0.090)}_{-0.057(0.082)}$ & $1.799^{+0.077(0.120)}_{-0.107(0.163)}$ & $0.291^{+0.034(0.057)}_{-0.028(0.046)}$ & $-$ & $0.346^{+0.177(0.264)}_{-0.157(0.283)}$ & $-1.160^{+0.682(0.900)}_{-4.313({\rm unc)}}$ \\
$0.712^{+0.048(0.086)}_{-0.060(0.092)}$ & $1.806^{+0.091(0.138)}_{-0.094(0.149)}$ & $0.291^{+0.030(0.051)}_{-0.029(0.044)}$ & $0.306^{+0.692(1.184)}_{-0.464(0.686)}$ & $0.252^{+0.150(0.277)}_{-0.173(0.251)}$ & $-2.244^{+1.774(2.285)}_{-2.436(2.864)}$ \\
\hline
\multicolumn{6}{c}{No LRG1} \\
\cline{1-6}
$0.713^{+0.062(0.100)}_{-0.047(0.077)}$ & $1.827^{+0.081(0.126)}_{-0.095(0.153)}$ & $0.294^{+0.030(0.054)}_{-0.029(0.043)}$ & $-$ & $0.556^{+0.165(0.284)}_{-0.173(0.440)}$ & $-5.673^{+5.197(5.469)}_{\rm\ unc}$ \\
$0.710^{+0.047(0.088)}_{-0.058(0.086)}$ & $1.860^{+0.074(0.122)}_{-0.084(0.142)}$ & $0.289^{+0.028(0.052)}_{-0.028(0.043)}$ & $0.715^{+0.684(1.693)}_{-0.957(1.192)}$ & $0.251^{+0.240(0.524)}_{-0.199(0.248)}$ & $-0.551^{+0.498(0.883)}_{-1.501(2.382)}$ \\
\hline
\end{tabular}
\caption{Best-fit parameters at $1$-$\sigma$ ($2$-$\sigma$) from the MCMC simulation of $E_p-E_{iso}$ correlation, calibrated with and without the DESI LRG1 data point, for the $\omega_0$CDM model with $\Omega_k= 0$ and $\Omega_k\neq 0$.}
\label{tab:DESIBAOwCDM}
\end{table*}

The simplest extension of the concordance paradigm, the $\omega_0$CDM model, is described by Eq.~\eqref{hrate} with
\begin{equation}
    G(z)\equiv (1+z)^{3(1+\omega_0)}\,,
\end{equation}
and reduces to the $\Lambda$CDM scenario when $\omega_0=-1$.

The results of our computations for such cosmological model are shown in Table~\ref{tab:DESIBAOwCDM}. As for the $\Lambda$CDM paradigm, the best-fit parameters for the $E_p-E_{iso}$ correlation are consistent with the results in the literature.

We compare our results on the cosmological parameters in both flat and non-flat scenarios with the outcomes of the \citet{2020A&A...641A...6P} denoted by the superscript $P$ and the SDSS denoted by the superscript $S$ \cite{2021PhRvD.103h3533A}.

The general conclusions for this cosmological model can be summarized as follows.
\begin{itemize}
    \item[-] All our results for the matter density $\Omega_m$ are in agreement at $1$-$\sigma$ with both the above-introduced values from Planck, $\Omega_m^P$ \cite{2020A&A...641A...6P}, and SDSS, $\Omega_m^S$ \cite{2021PhRvD.103h3533A}, in both flat and non-flat scenarios. The only exception is the value got from the flat $\omega_0$CDM model without considering the LRG1 point, which is anomalously high and consistent with $\Omega_m^P$ and $\Omega_m^S$ only at $2$-$\sigma$.
    \item[-] All the estimates of the barotropic factor $\omega_0$ have enormous attached errors, which make them all consistent at $1$-$\sigma$ with the expectation of the $\Lambda$CDM model, namely  $\omega_0=-1$.
    \item[-] Focusing on the curvature parameter, we can compare our results with the above-discussed estimates from Planck $\Omega_k^P=-0.011^{+0.013}_{-0.012}$ \cite{2020A&A...641A...6P} and SDSS $\Omega_k^S = 0.079^{+0.083}_{-0.100}$ \cite{2021PhRvD.103h3533A}. Both considering or excluding the LRG1 point, the analyses lead to positive and non-negligible curvature parameters. However, the attached errors are very large and, eventually, our results turn out to be consistent with the above $\Omega_k^P$ and $\Omega_k^S$ and, thus with the flat case $\Omega_k\equiv0$.
\end{itemize}

Unlike previous findings for the $\Lambda$CDM model, the $h_0$--$\Omega_m$ degeneracy, due to the different values of $h_0$ in Eqs.~\eqref{h0142}--\eqref{h0152} got by calibrating with or without the LRG1 point, is evident only for flat scenarios.
When considering non-flat models, both estimates tend to a common value of $\Omega_m\simeq 0.25$, which is still consistent with both Planck and SDSS results.

As mentioned before, the flat $\omega_0$CDM model without the LRG1 point provides a very high value of the mass, consistent with $\Omega_m\simeq0.55$. This result seems in line with the general trend that GRBs provide high-mass results and also with the recent results got from quasars \cite{risaliti2019cosmological}.

 \begin{table*}
\centering
\setlength{\tabcolsep}{0.25em}
\renewcommand{\arraystretch}{1.7}
\begin{tabular}{ccccccc}
\hline
$a$ & $b$ & $\sigma$ & $\Omega_k$ & $\Omega_m$ & $\omega_0$ & $\omega_1$\\
\hline
\multicolumn{7}{c}{With LRG1} \\
\cline{1-7}
$0.726^{+0.045(0.084)}_{-0.065(0.093)}$ & $1.785^{+0.092(0.135)}_{-0.096(0.153)}$ & $0.297^{+0.027(0.048)}_{-0.035(0.049)}$ & $-$ & $0.378^{+0.128(0.244)}_{-0.122(0.266)}$ & $-3.609^{+2.784(3.391)}_{-2.212(2.390)}$ &
$0.007^{+1.595(2.742)}_{-1.797(3.273)}$ \\
$0.709^{+0.048(0.083)}_{-0.057(0.084)}$ & $1.816^{+0.070(0.129)}_{-0.110(0.166)}$ & $0.291^{+0.030(0.052)}_{-0.029(0.047)}$ & $0.386^{+0.434(0.786)}_{-0.373(0.607)}$ & $0.229^{+0.180(0.276)}_{-0.132(0.219)}$ & $-5.650^{+3.002(4.169)}_{\rm\ unc}$  &
$2.157^{+3.841({\rm unc})}_{-2.818(4.343)}$ \\
\hline
\multicolumn{7}{c}{No LRG1} \\
\cline{1-7}
$0.704^{+0.066(0.104)}_{-0.043(0.078)}$ & $1.840^{+0.080(0.129)}_{-0.102(0.160)}$ & $0.290^{+0.033(0.053)}_{-0.028(0.043)}$ & $-$ & $0.505^{+0.200(0.355)}_{-0.229(0.426)}$ & $-2.420^{+3.459(3.793)}_{-2.209(3.524)}$ &
$-4.041^{+3.851(6.842)}_{-3.928(5.440)}$ \\
$0.710^{+0.054(0.096)}_{-0.054(0.089)}$ & $1.834^{+0.083(0.135)}_{-0.092(0.148)}$ & $0.291^{+0.030(0.054)}_{-0.030(0.044)}$ & $0.197^{+0.469(0.688)}_{-0.456(0.606)}$ & $0.269^{+0.197(0.351)}_{-0.184(0.269)}$ & $-0.917^{+0.722(0.884)}_{-4.196(4.667)}$ &
$-0.282^{+1.656(2.462)}_{-1.424(2.574)}$ \\
\hline
\end{tabular}
\caption{Best-fit parameters at $1$-$\sigma$ ($2$-$\sigma$) from the MCMC simulation of $E_p-E_{iso}$ correlation, calibrated with and without the DESI LRG1 data point, for the $\omega_0\omega_1$CDM model with $\Omega_k= 0$ and $\Omega_k\neq 0$.}
\label{tab:DESIBAOCPL}
\end{table*}

\subsection{The $\omega_0\omega_1$CDM model}

In the $\omega_0\omega_1$CDM model, first introduced by \citet{2001IJMPD..10..213C} and then by \citet{2003PhRvL..90i1301L}, the barotropic factor evolves with $z$, \textit{i.e.} $\omega(z) = \omega_0+\omega_1 z(1+z)^{-1}$. For this model, it turns out to be
\begin{equation}
   G(z)\equiv (1+z)^{3(1+\omega_0+\omega_1)}\exp\left(-\frac{3\omega_1z}{1+z}\right).
\end{equation}

The results of our MCMC simulation for this cosmological model are shown in Table~\ref{tab:DESIBAOCPL}. Also in this case the best-fit parameters of the $E_p-E_{iso}$ correlation are in agreement with results found in the literature.

As for the previous two cases we compare the cosmological parameters inferred from our MCMC simulations with the values of the SDSS and Planck, labelled with the superscript $S$ and $P$, respectively.

\begin{itemize}
    \item[-] The matter density $\Omega_m$ values we found agree at $1$-$\sigma$ with the values from Planck, $\Omega_m^P$ \cite{2020A&A...641A...6P}, and SDSS, $\Omega_m^S$ \cite{2021PhRvD.103h3533A}, in both flat and non-flat scenarios.
    Remarkably, the value from the flat scenario without considering the LRG1 point, provides an anomalously high value of $\Omega_m$, in analogy to the case of the flat $\omega_0$CDM model.
    \item[-] The values of $\omega_0$ and $\omega_1$ have very large attached error bars, which make them compatible with the expectation of the concordance paradigm, \textit{i.e.} $\omega_0 = -1$ and $\omega_1 = 0$ at $1$-$\sigma$.
    There only one exception, given by the non-flat scenario with the whole DESI data set, where $\omega_0$ in inconsistent beyond $2$-$\sigma$ with the value $\omega_0 = -1$.
    \item[-] By comparison with the above-discussed estimates from Planck $\Omega_k^P=-0.011^{+0.013}_{-0.012}$ \cite{2020A&A...641A...6P} and SDSS $\Omega_k^S = 0.079^{+0.083}_{-0.100}$ \cite{2021PhRvD.103h3533A}, considering or excluding the LRG1 point lead to positive and non-negligible curvature parameters. Also for this model, the attached errors are such that our $\Omega_k$ turn out to be consistent with the above $\Omega_k^P$ and $\Omega_k^S$ and, thus with the flat case $\Omega_k\equiv0$.
\end{itemize}

Even in this case, the $h_0$--$\Omega_m$ degeneracy affects the estimates of $\Omega_m$, as they are intertwined with the values of $h_0$ in Eqs.~\eqref{h0142}--\eqref{h0152}.
Like the case of the $\omega_0$CDM model, in the non-flat scenario, considering or excluding the LRG1 point provides estimates that tend to $\Omega_m\simeq 0.23$--$0.27$, which is still consistent with both Planck and SDSS results.

\subsection{Model selection criteria and statistical analysis}

 \begin{table*}
\centering
\setlength{\tabcolsep}{.8em}
\renewcommand{\arraystretch}{1.7}
\begin{tabular}{l|lcccccccccc}
\hline
   & & $-\ln \mathcal L_m$ & \text{AIC} & AICc & BIC & DIC & $\Delta$AIC & $\Delta$AICc & $\Delta$BIC & $\Delta$DIC\\
\hline
\multirow{6}{*}{\rm With\ LRG1} &
$\Lambda$CDM ($\Omega_k=0$) & $189.38$ & $387$ & $387$ & $398$ & $387$ & $0$ & $0$ & $0$ & $0$ \\
&
$\Lambda$CDM ($\Omega_k\neq0$) & $189.20$ & $389$ & $389$ & $402$ & $389$ & $2$ & $2$ & $4$ & $2$ \\
&
$\omega_0$CDM ($\Omega_k=0$) & $189.36$ & $389$ & $389$ & $403$ & $390$ & $2$ & $2$ & $5$ & $3$\\
&
$\omega_0$CDM ($\Omega_k\neq0$) & $188.89$ & $390$ & $390$ & $406$ & $389$ & $3$ & $3$ & $8$ & $2$ \\
&
$\omega_0\omega_1$CDM ($\Omega_k=0$) &   $189.47$ & $391$ & $392$ & $408$ & $389$ & $4$ & $5$ & $10$ & $2$\\
&
$\omega_0\omega_1$CDM ($\Omega_k\neq0$) &  $188.91$ & $392$ & $393$ & $411$ & $389$ & $5$ & $6$ & $13$ & $2$ \\
\hline
\multirow{6}{*}{\rm No\ LRG1} &
$\Lambda$CDM ($\Omega_k=0$) & $189.55$ & $387$ & $387$ & $398$ & $388$ & $0$ & $0$ & $0$ & $0$ \\
&
$\Lambda$CDM ($\Omega_k\neq0$) & $189.09$ & $388$ & $389$ & $402$ & $389$ & $1$ & $2$ & $4$ & $1$ \\
&
$\omega_0$CDM ($\Omega_k=0$) & $189.10$  & $388$ & $389$ & $402$ & $389$ & $1$ & $2$ & $4$ & $1$\\
&
$\omega_0$CDM ($\Omega_k\neq0$) & $189.09$ & $390$ & $391$ & $407$ & $389$ & $3$ & $4$ & $9$ & $1$ \\
&
$\omega_0\omega_1$CDM ($\Omega_k=0$) &  $189.08$ & $390$ & $391$ & $407$ & $391$ & $3$ & $4$ & $13$ & $3$\\
&
$\omega_0\omega_1$CDM ($\Omega_k\neq0$) & $188.94$ & $392$ & $393$ & $411$ & $389$ & $5$ & $6$ & $13$ & $1$ \\
\hline
\end{tabular}
\caption{Comparison between flat and non-flat $\Lambda$CDM, $\omega_0$CDM and $\omega_0\omega_1$CDM models using criteria of model selections.}
\label{tab:modelsel}
\end{table*}

The DESI collaboration claim that the new released BAO data favour the $\omega_0\omega_1$CDM model \cite{2024arXiv240403002D}.

To verify this claim through our MCMC simulations based on GRB data calibrated via OHD and DESI-BAO data sets, we apply the following model selection criteria:
\begin{enumerate}
    \item [-] the Akaike information criterion (AIC) \cite{akaike2011akaike, Akaike:1998zah, 2007MNRAS.377L..74L}
    \begin{equation}\label{AIC}
        \text{AIC}\equiv -2\ln\mathcal{L}_{m}+2d\,;
    \end{equation}
    \item [-] the corrected AIC (AICc)
    \begin{equation}
        \text{AICc}\equiv \text{AIC}+\frac{2d(d+1)}{N-d-1}\,;
    \end{equation}
    \item [-] the Bayesian information criterium (BIC) \cite{2007MNRAS.377L..74L}
    \begin{equation}
        \text{BIC}\equiv-2\ln\mathcal{L}_{m}+2d\ln(N)\,;
    \end{equation}
    \item [-] the deviance information criterion (DIC) \cite{2006PhRvD..74b3503K, 2007MNRAS.377L..74L, 10.1111/1467-9868.00353}
    \begin{equation}\label{DIC}
        \text{DIC}\equiv 2\left(\langle-2\ln\mathcal{L}\rangle+2\ln\mathcal{L}_{m}\right)-2\ln\mathcal{L}_{m}\,.
    \end{equation}
\end{enumerate}
In Eqs.~\eqref{AIC}--\eqref{DIC} the maximum log-likelihood $\ln\mathcal{L}_{m}$ is considered, $d$ are the number of free parameters in the model, $N$ is the number of datapoints used in our computations and $\langle-2\ln\mathcal{L}\rangle$ is the average over the posterior distribution. It is worth to stress that the DIC criterium, on the contrary of the AIC, AICc and BIC criteria, does discard unconstrained parameters \cite{2007MNRAS.377L..74L}.

In our analyses we have $N=118$ GRB data points of the $E_p-E_{iso}$ correlation, whereas the number of parameters $d$ vary from model to model, \textit{i.e.} $d=4$ for the flat $\Lambda$CDM, $d=5$ for the non-flat $\Lambda$CDM and flat $\omega_0$CDM, $d = 6$ for the non-flat $\omega_0$CDM and flat $\omega_0\omega_1$CDM and $d = 7$ for the non-flat $\omega_0\omega_1$CDM.

The most-suited model is the one providing the the lowest selection criteria, \textit{i.e.} $X_0$, and the lowest difference $\Delta X=X_j-X_0$, where $X_j = \text{AIC, AICc, BIC, DIC}$.

From Table~\ref{tab:modelsel}, the preferred model is indeed the flat $\Lambda$CDM model, independently on the calibrations with or without the LRG1 point (see also Table~\ref{tab:DESIBAOLCDM}).
A part for the $\Delta$DIC criterion, all the other criteria seem to indicate that, for both calibrations: i) the non-flat $\Lambda$CDM and the flat $\omega_0$CDM models (both with $p=5$) are weakly excluded, and ii) the remaining models are more and more disfavoured as the number of parameters increases (see also Tables~\ref{tab:DESIBAOwCDM}--\ref{tab:DESIBAOCPL}).

\section{Outlooks and perspectives}\label{conc}

Recently, \citet{2024arXiv240403002D} found intersting results that are already a hot topic of discussion among the scientific community \cite{2024arXiv240412068C, 2024arXiv240407070L, 2024arXiv240408633C, 2024arXiv240504216C, 2024arXiv240513588L, 2024arXiv240418579W, 2024arXiv240500502P}.

In this work, we calibrated the $E_p-E_{iso}$ correlation for GRBs using B\'ezier interpolations of both OHD and the DESI 2024 data, with the aim of circumventing the \textit{circularity problem}.
In particular, we considered:
\begin{enumerate}
    \item [-] the entire DESI data sample,
    \item [-] the DESI data sample without the LRG1 point placed at $z_{eff} = 0.51$, in analogy to the analyses reported in Refs. \cite{2024arXiv240408633C, 2024arXiv240412068C}.
\end{enumerate}

Afterwards, we used the calibrated correlation to test three different cosmological models, namely,
\begin{itemize}
    \item[-] the $\Lambda$CDM, or concordance paradigm,
    \item[-] the $\omega_0$CDM model,
    \item[-] the $\omega_0\omega_1$CDM scenario.
\end{itemize}
In particular, the latter appears favoured by DESI observations \cite{2024arXiv240403002D}, showing that a possible evolving dark energy can frame the large scale dynamics. We have checked if even GRBs may lean in this direction, by working out both spatially-flat and non-flat cosmologies.

In so doing, we employed MCMC simulations using the Metropolis-Hastings algorithm \cite{1953JChPh..21.1087M, 1970Bimka..57...97H}. Then, we compared our evaluated matter density $\Omega_m$ and curvature $\Omega_k$ parameters with the ones from the \citet{2020A&A...641A...6P} and the SDSS \cite{2021PhRvD.103h3533A}.

We found that, for the $\Lambda$CDM paradigm, $\Omega_m$ is mostly consistent at $1$-$\sigma$ with Planck \cite{2020A&A...641A...6P} and SDSS \cite{2021PhRvD.103h3533A}, with the only exception of the non-flat case, calibrated with the whole DESI sample, where we got $\Omega_m\simeq0.2$ that agrees only at $2$-$\sigma$ with Planck \cite{2020A&A...641A...6P} and at $1$-$\sigma$ with SDSS \cite{2021PhRvD.103h3533A}.
Concerning the curvature parameter $\Omega_k$, our estimates have always large attached errors, making them consistent at $1$-$\sigma$ with both the values from the SDSS \cite{2021PhRvD.103h3533A} and Planck \cite{2020A&A...641A...6P} and, thus, with the flat scenario $\Omega_k\equiv0$.

Concerning the $\omega_0$CDM model, our results for the matter density $\Omega_m$ are in agreement at $1$-$\sigma$ with the values from Planck \cite{2020A&A...641A...6P} and SDSS \cite{2021PhRvD.103h3533A}, in both flat and non-flat scenarios. Only the value from the flat $\omega_0$CDM model obtained by excluding the LRG1 point is anomalously high and consistent with $\Omega_m^P$ and $\Omega_m^S$ only at $2$-$\sigma$.
On the other hand, our outcomes on the barotropic factor $\omega_0$ and the curvature parameter $\Omega_k$ have very large attached errors, which make all the results always consistent at $1$-$\sigma$ with both Planck and SDSS values and, thus, with the concordance paradigm, \textbf{i.e.} $\omega_0\equiv-1$ and $\Omega_k\equiv0$.

Finally, for the $\omega_0\omega_1$CDM our values of $\Omega_m$ agree at $1$-$\sigma$ with the values from Planck \cite{2020A&A...641A...6P} and SDSS \cite{2021PhRvD.103h3533A}, in both flat and non-flat cases. Remarkably, the value from the flat scenario, without considering the LRG1 point, provides a very large value $\Omega_m\simeq0.5$, in analogy to the case of the flat $\omega_0$CDM model.
The values of $\omega_0$ and $\omega_1$ in view of the large errors, are mostly compatible with the concordance paradigm, \textit{i.e.} $\omega_0 = -1$ and $\omega_1 = 0$ at $1$-$\sigma$. Only in the non-flat scenario, with the whole DESI data set, $\omega_0$ in inconsistent beyond $2$-$\sigma$ with $\omega_0 = -1$.
In all our analyses, the estimates on $\Omega_k$ have large attached errors such that they turn out to be consistent with the values from Planck \cite{2020A&A...641A...6P} and SDSS \cite{2021PhRvD.103h3533A} in both flat and non-flat cases and,thus, with the flat case $\Omega_k\equiv0$.

In all three cosmological models, we noticed that the higher values of $\Omega_m$ are mostly inferred from the DESI catalog without LRG1. This is a direct consequence of the well-known $h_0$--$\Omega_m$ degeneracy \cite{1997ApJ...488....1Z, 2001ApJ...549..669H, 2002MNRAS.337.1068P}. Further, we underline that it is also well-established that GRBs boost $\Omega_m$ to higher bounds with respect to those extracted by other probes \cite{sym12071118, amati2019addressing}, though this is not always the case in our analyses.
Finally, we notice that in general, when curvature is introduced, the mass bounds tend to decrease and to have similar values between the calibrations with or without the LRG1 data point.

To conclude our discussion, we checked the preferred cosmological models by using the widely used model selection criteria. We found that, besides for the DIC, the most favored cosmological model remains the $\Lambda$CDM paradigm. We also found a weak preference for the non-flat $\Lambda$CDM and the flat $\omega_0$CDM models.
These results seems to conclude that GRBs do not exclude the presence of a (small) curvature or an evolving dark energy term with $\omega_0\neq-1$.

Future works will be focused on next DESI data release to check if the preference towards the $\omega_0\omega_1$CDM still persists. Moreover, it would be interesting to check if our results persist even when adopting additional GRB correlation functions, \textit{e.g.} other prompt emission correlations such as the $L_p-E_p$ (or Yonetoku) or prompt emission and afterglow correlations such as the $L_0-E_p-T$ (or Combo) \cite{2004ApJ...609..935Y, 2015A&A...582A.115I}.

\section*{Acknowledgements}

OL acknowledges Alejandro Aviles and Eoin O Colgain for private discussions related to the subject of this work. ACA  acknowledge the Istituto Nazionale di Fisica Nucleare (INFN) Sez. di Napoli, Iniziativa Specifica QGSKY.

%

\newpage

\appendix

\begin{widetext}

\section{Contour plots and best-fit B\'ezier coefficients}\label{appendix1}

In Tables~\ref{tab:DESIBAOrdvar}--\ref{tab:DESIBAOrdvarwithout051} we list the best-fit B\'ezier coefficients and the corresponding values of the log-likelihood functions, obtained by considering $r_d$ within the interval $r_d\in [138, 156]$ and $r_d=147.09$ from \citet{2020A&A...641A...6P}.
\begin{table*}
\centering
\setlength{\tabcolsep}{0.25em}
\renewcommand{\arraystretch}{1.7}
\begin{tabular}{lcccccccc}
\hline
$r_d$     & $\alpha_0\equiv h_0$ & $\alpha_1$ & $\alpha_2$ & $\beta_1$ & $\beta_2$ & $\beta_3$ & $-\ln\mathcal{L}$\\
\hline
$138$ & $0.726^{+0.025(0.049)}_{-0.023(0.049)}$ & $1.103^{+0.016(0.035)}_{-0.018(0.037)}$ & $2.140^{+0.009(0.017)}_{-0.006(0.014)}$ & $-1.506^{+0.804(1.764)}_{-0.646(1.457)}$ & $42.87^{+3.61(8.09)}_{-4.78(9.45)}$ & $334.462^{+12.712(28.270)}_{-11.753(23.621)}$ & $157.91$\\
$140$ & $0.723^{+0.027(0.051)}_{-0.017(0.045)}$ & $1.087^{+0.012(0.033)}_{-0.024(0.042)}$ & $2.110^{+0.006(0.014)}_{-0.008(0.017)}$ & $-1.563^{+0.971(1.858)}_{-0.527(1.393)}$ & $44.35^{+2.88(7.49)}_{-5.52(9.71)}$ & $343.895^{+14.949(29.349)}_{-9.680(23.617)}$ & $156.28$\\
$142$ & $0.729^{+0.019(0.046)}_{-0.027(0.054)}$ & $1.061^{+0.018(0.038)}_{-0.016(0.034)}$ & $2.077^{+0.009(0.016)}_{-0.007(0.014)}$ & $-1.493^{+1.000(1.949)}_{-0.637(1.511)}$ & $44.85^{+3.67(8.60)}_{-5.00(9.27)}$ & $357.336^{+13.164(26.698)}_{-14.957(28.398)}$ & $154.95$\\
$144$ & $0.723^{+0.028(0.050)}_{-0.024(0.047)}$ & $1.042^{+0.020(0.038)}_{-0.016(0.037)}$ & $2.049^{+0.006(0.014)}_{-0.009(0.017)}$ & $-1.221^{+0.846(1.809)}_{-0.912(1.793)}$ & $45.09^{+4.70(9.28)}_{-4.07(9.52)}$ & $368.545^{+10.540(27.156)}_{-17.251(29.436)}$ & $153.91$\\
$146$ & $0.726^{+0.022(0.047)}_{-0.028(0.052)}$ & $1.025^{+0.018(0.036)}_{-0.017(0.035)}$ & $2.020^{+0.006(0.014)}_{-0.008(0.016)}$ & $-1.138^{+0.758(1.750)}_{-1.002(1.881)}$ & $45.56^{+5.44(10.45)}_{-4.38(9.09)}$ & $377.201^{+15.289(29.030)}_{-14.353(28.560)}$ & $153.13$\\
$148$  & $0.717^{+0.028(0.053)}_{-0.019(0.045)}$ & $1.008^{+0.016(0.037)}_{-0.017(0.035)}$ & $1.990^{+0.007(0.014)}_{-0.008(0.015)}$ & $-1.438^{+1.067(2.196)}_{-0.703(1.660)}$ & $48.58^{+2.86(8.94)}_{-5.39(11.06)}$ & $384.961^{+16.327(32.384)}_{-11.601(26.403)}$ & $152.54$\\
$150$ & $0.722^{+0.023(0.045)}_{-0.026(0.051)}$ & $0.989^{+0.018(0.037)}_{-0.014(0.033)}$ & $1.961^{+0.008(0.016)}_{-0.006(0.014)}$ & $-1.363^{+1.205(2.151)}_{-0.723(1.724)}$ & $49.37^{+4.25(9.38)}_{-5.58(11.47)}$ & $394.598^{+16.610(34.998)}_{-10.336(26.492)}$ & $152.22$\\
$152$ & $0.718^{+0.022(0.048)}_{-0.020(0.046)}$ & $0.976^{+0.016(0.033)}_{-0.021(0.036)}$ & $1.936^{+0.006(0.012)}_{-0.007(0.016)}$ & $-1.333^{+1.262(2.400)}_{-0.640(1.742)}$ & $51.10^{+3.11(8.84)}_{-6.45(11.87)}$ & $404.383^{+18.090(37.317)}_{-9.648(25.727)}$ & $152.11$\\
$154$ & $0.723^{+0.018(0.042)}_{-0.029(0.054)}$ & $0.952^{+0.024(0.043)}_{-0.011(0.029)}$ & $1.908^{+0.009(0.015)}_{-0.007(0.013)}$ & $-0.897^{+0.896(2.020)}_{-1.177(2.241)}$ & $49.61^{+7.01(11.73)}_{-4.28(10.45)}$ & $424.307^{+10.025(29.412)}_{-19.833(35.920)}$ & $152.24$\\
$156$ & $0.712^{+0.027(0.049)}_{-0.020(0.044)}$ & $0.944^{+0.014(0.035)}_{-0.017(0.035)}$ & $1.882^{+0.008(0.014)}_{-0.005(0.015)}$ & $-1.205^{+1.251(2.310)}_{-0.811(1.858)}$ & $52.44^{+4.50(10.09)}_{-6.19(11.63)}$ & $430.247^{+18.636(35.135)}_{-12.474(31.325)}$ & $152.48$\\
\hline
$147.09$ & $0.721^{+0.025(0.051)}_{-0.021(0.049)}$ & $1.017^{+0.017(0.035)}_{-0.016(0.036)}$ & $2.003^{+0.007(0.014)}_{-0.007(0.015)}$ & $-1.226^{+0.996(1.885)}_{-0.847(1.841)}$ & $47.92^{+3.31(9.02)}_{-5.02(10.79)}$ & $378.319^{+18.837(33.688)}_{-10.304(26.158)}$ & $152.78$
\\
\hline
\end{tabular}
\caption{B\'ezier coefficients with associated errors at $1$--$\sigma$ ($2$--$\sigma$) for values of the sound horizon in the interval $r_d\in[138, 156]$ Mpc and considering the complete DESI catalog.}
\label{tab:DESIBAOrdvar}
\end{table*}
\begin{table*}
\centering
\setlength{\tabcolsep}{0.25em}
\renewcommand{\arraystretch}{1.7}
\begin{tabular}{lcccccccc}
\hline
$r_d$     & $\alpha_0\equiv h_0$ & $\alpha_1$ & $\alpha_2$ & $\beta_1$ & $\beta_2$ & $\beta_3$ & $-\ln\mathcal{L}$\\
\hline
$138$ & $0.641^{+0.020(0.054)}_{-0.034(0.070)}$ & $1.129^{+0.022(0.043)}_{-0.017(0.034)}$ & $2.152^{+0.009(0.018)}_{-0.006(0.015)}$ & $-1.941^{+0.768(1.731)}_{-0.605(1.529)}$ & $40.92^{+4.35(8.72)}_{-4.49(9.74)}$ & $341.727^{+10.847(26.337)}_{-13.197(27.595)}$ & $143.56$\\
$140$ & $0.638^{+0.029(0.060)}_{-0.030(0.062)}$ & $1.110^{+0.021(0.041)}_{-0.019(0.038)}$ & $2.120^{+0.009(0.017)}_{-0.007(0.015)}$ & $-1.942^{+0.900(1.773)}_{-0.708(1.382)}$ & $42.60^{+3.79(9.05)}_{-5.52(10.26)}$ & $350.740^{+11.989(28.583)}_{-11.030(27.241)}$ & $143.32$\\
$142$ & $0.646^{+0.023(0.054)}_{-0.036(0.068)}$ & $1.085^{+0.025(0.044)}_{-0.014(0.033)}$ & $2.088^{+0.009(0.017)}_{-0.007(0.015)}$ & $-1.586^{+0.575(1.538)}_{-1.538(1.804)}$ & $42.19^{+4.82(9.64)}_{-4.85(9.51)}$ & $360.954^{+13.294(29.057)}_{-11.321(28.320)}$ & $143.30$\\
$144$ & $0.641^{+0.033(0.062)}_{-0.030(0.058)}$ & $1.074^{+0.017(0.036)}_{-0.021(0.040)}$ & $2.059^{+0.007(0.015)}_{-0.008(0.017)}$ & $-1.661^{+0.794(1.757)}_{-0.943(1.776)}$ & $43.04^{+5.75(11.09)}_{-4.05(9.37)}$ & $375.767^{+9.876(25.815)}_{-18.496(33.860)}$ & $143.43$\\
$146$ & $0.654^{+0.021(0.054)}_{-0.034(0.070)}$ & $1.049^{+0.022(0.042)}_{-0.016(0.037)}$ & $2.027^{+0.009(0.017)}_{-0.007(0.014)}$ & $-1.628^{+0.762(1.873)}_{-1.044(1.855)}$ & $43.71^{+6.33(11.64)}_{-4.39(9.70)}$ & $385.087^{+12.863(28.370)}_{-18.376(33.128)}$ & $143.78$ \\
$148$  & $0.654^{+0.026(0.058)}_{-0.033(0.064)}$ & $1.027^{+0.022(0.042)}_{-0.013(0.033)}$ & $1.998^{+0.009(0.018)}_{-0.005(0.014)}$ & $-1.511^{+0.798(1.766)}_{-1.058(1.962)}$ & $45.75^{+4.83(10.77)}_{-5.09(10.69)}$ & $393.517^{+13.946(30.478)}_{-16.658(31.655)}$ & $144.24$\\
$150$ & $0.643^{+0.035(0.071)}_{-0.020(0.051)}$ & $1.020^{+0.012(0.033)}_{-0.025(0.042)}$ & $1.973^{+0.006(0.014)}_{-0.009(0.017)}$ & $-1.554^{+0.936(1.932)}_{-0.990(1.886)}$ & $46.44^{+5.58(11.01)}_{-5.23(10.87)}$ & $402.569^{+17.031(33.891)}_{-12.855(31.478)}$ & $144.84$\\
$152$ & $0.648^{+0.039(0.066)}_{-0.024(0.051)}$ & $0.998^{+0.017(0.034)}_{-0.019(0.041)}$ & $1.943^{+1.089(2.253)}_{-7.762(1.759)}$ & $-1.711^{+1.089(2.253)}_{-0.776(1.759)}$ & $49.15^{+4.12(10.08)}_{-6.98(12.84)}$ & $413.515^{+17.967(35.791)}_{-14.416(30.403)}$ & $145.63$\\
$154$ & $0.654^{+0.030(0.062)}_{-0.026(0.055)}$ & $0.982^{+0.014(0.033)}_{-0.019(0.039)}$ & $1.917^{+0.007(0.014)}_{-0.008(0.014)}$ & $-1.424^{+0.958(2.046)}_{-0.985(2.061)}$ & $48.63^{+5.76(12.34)}_{-5.57(11.37)}$ & $423.788^{+18.572(35.449)}_{-14.375(30.283)}$ & $146.46$\\
$156$ & $0.666^{+0.018(0.051)}_{-0.037(0.051)}$ & $0.959^{+0.021(0.041)}_{-0.014(0.033)}$ & $1.890^{+0.008(0.016)}_{-0.006(0.015)}$ & $-1.057^{+0.735(1.908)}_{-1.264(2.369)}$ & $49.23^{+6.54(12.18)}_{-5.43(11.54)}$ & $437.141^{+16.539(35.622)}_{-18.871(34.670)}$ & $147.50$\\
\hline
$147.09$ & $0.642^{+0.034(0.064)}_{-0.022(0.054)}$ & $1.043^{+0.017(0.037)}_{-0.019(0.040)}$ & $2.013^{+0.007(0.015)}_{-0.008(0.016)}$ & $-1.575^{+0.756(1.817)}_{-0.854(1.817)}$ & $44.27^{+5.78(11.64)}_{-3.66(9.45)}$ & $390.622^{+11.892(27.124)}_{-16.897(34.059)}$ & $143.97$
\\
\hline
\end{tabular}
\caption{B\'ezier coefficients with associated errors at $1$--$\sigma$ ($2$--$\sigma$) for values of the sound horizon in the interval $r_d\in[138, 156]$ Mpc and considering the DESI catalog without the LRG1 data point.}
\label{tab:DESIBAOrdvarwithout051}
\end{table*}

Here we report the contour plots of the best-fit parameters of the $E_p-E_{iso}$ correlation and for the flat and non-flat $\Lambda$CDM, $\omega_0$CDM and $\omega_0\omega_1$CDM.

Specifically, Fig.~\ref{fig:1} displays the contour plot for the flat concordance paradigm while Fig.~\ref{fig:2} considers its non-flat extension. Then, Figs.~\ref{fig:3}-\ref{fig:4} show the contour plot of the flat and non-flat $\omega_0$CDM while Figs.~\ref{fig:5}-\ref{fig:6} are portraited the flat and non-flat scenario of the $\omega_0\omega_1$CDM model.

\begin{figure*}
\centering
{\hfill
\includegraphics[width=0.48\hsize,clip]{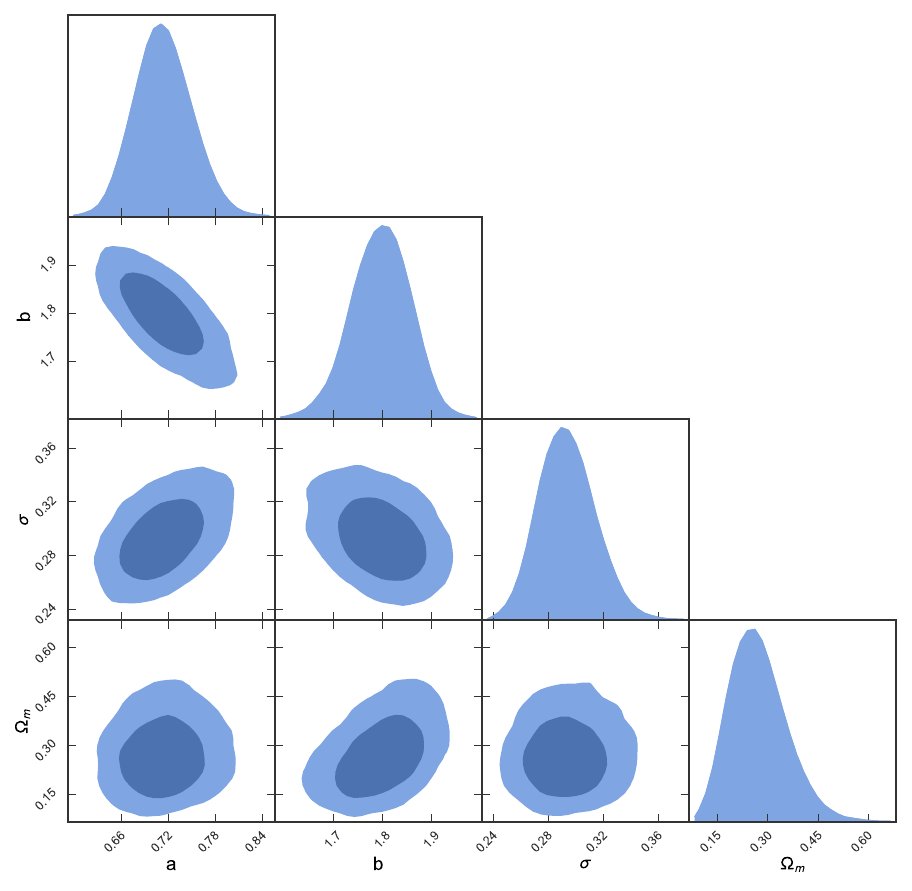}
\hfill
\includegraphics[width=0.48\hsize,clip]{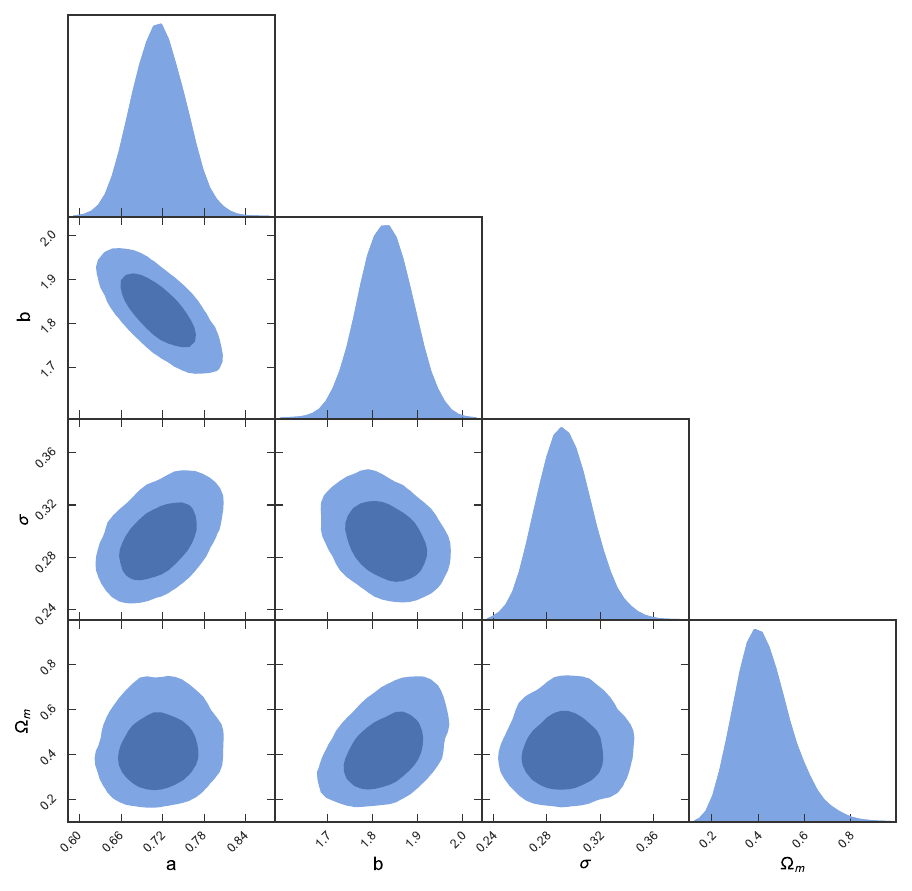}
\hfill}
\caption{Contour plots of the best-fit parameters for the $E_p-E_{iso}$ correlation and flat $\Lambda$CDM model without LRG1 (right) and with LRG1 (left).}
\label{fig:2}
\end{figure*}

\begin{figure*}
\centering
{\hfill
\includegraphics[width=0.48\hsize,clip]{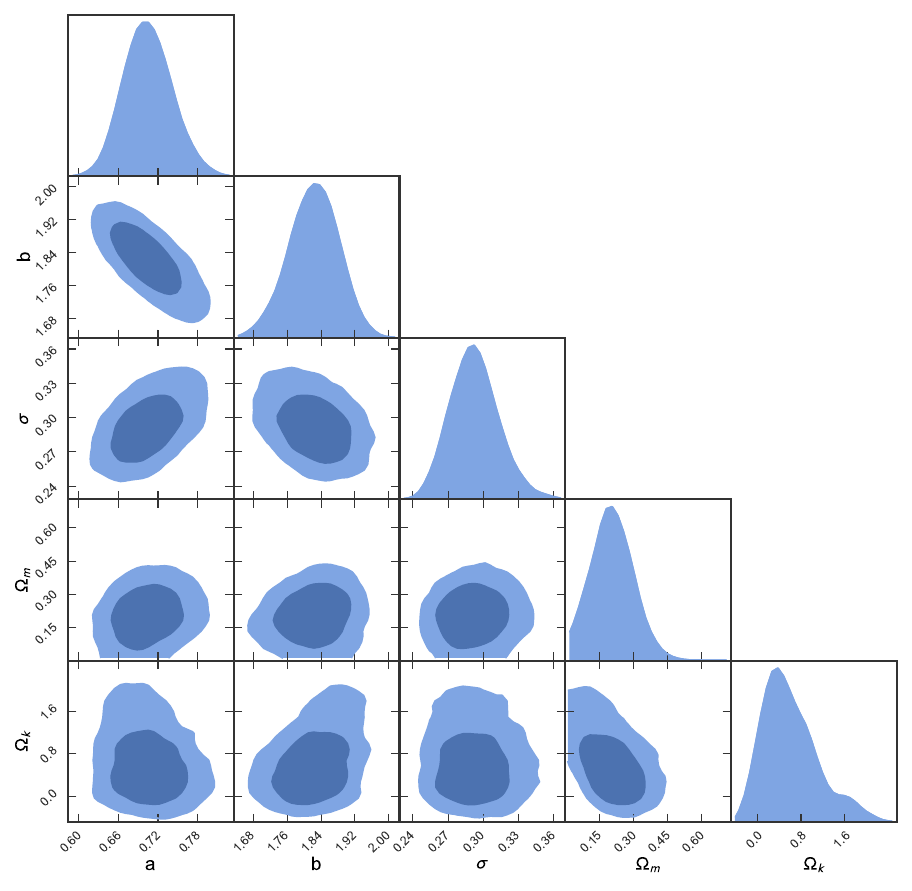}
\hfill
\includegraphics[width=0.48\hsize,clip]{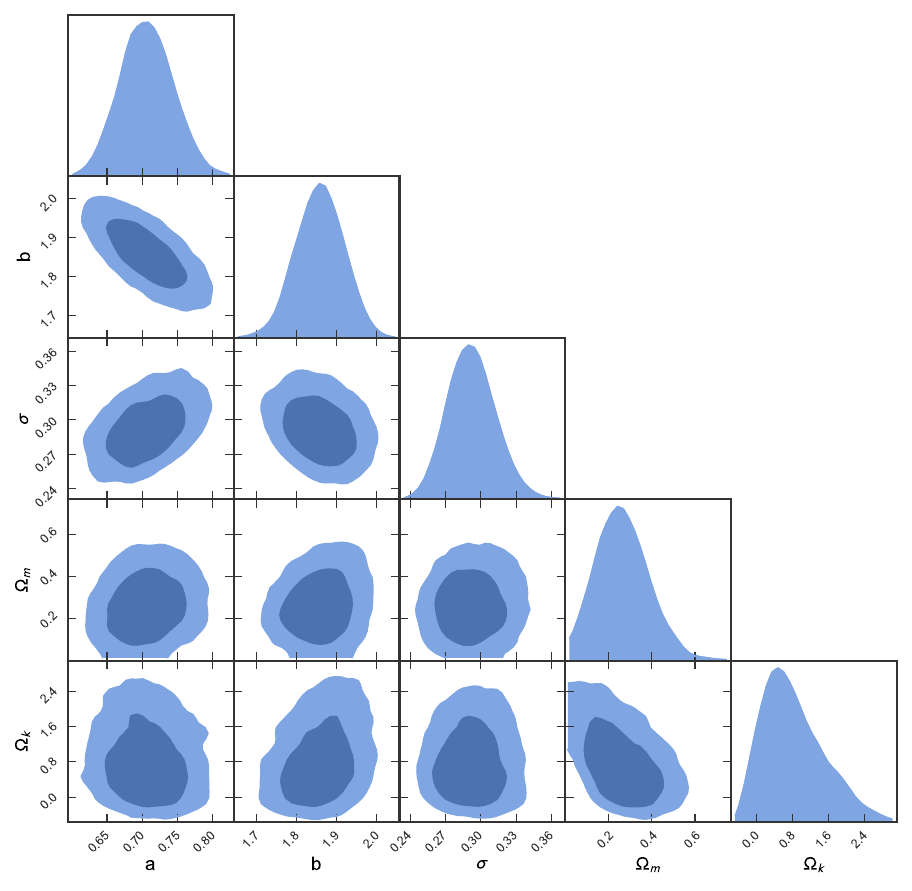}
\hfill}
\caption{Contour plots of the best-fit parameters for the $E_p-E_{iso}$ correlation and non-flat $\Lambda$CDM model without LRG1 (right) and with LRG1 (left).}
\label{fig:3}
\end{figure*}

\begin{figure*}
\centering
{\hfill
\includegraphics[width=0.48\hsize,clip]{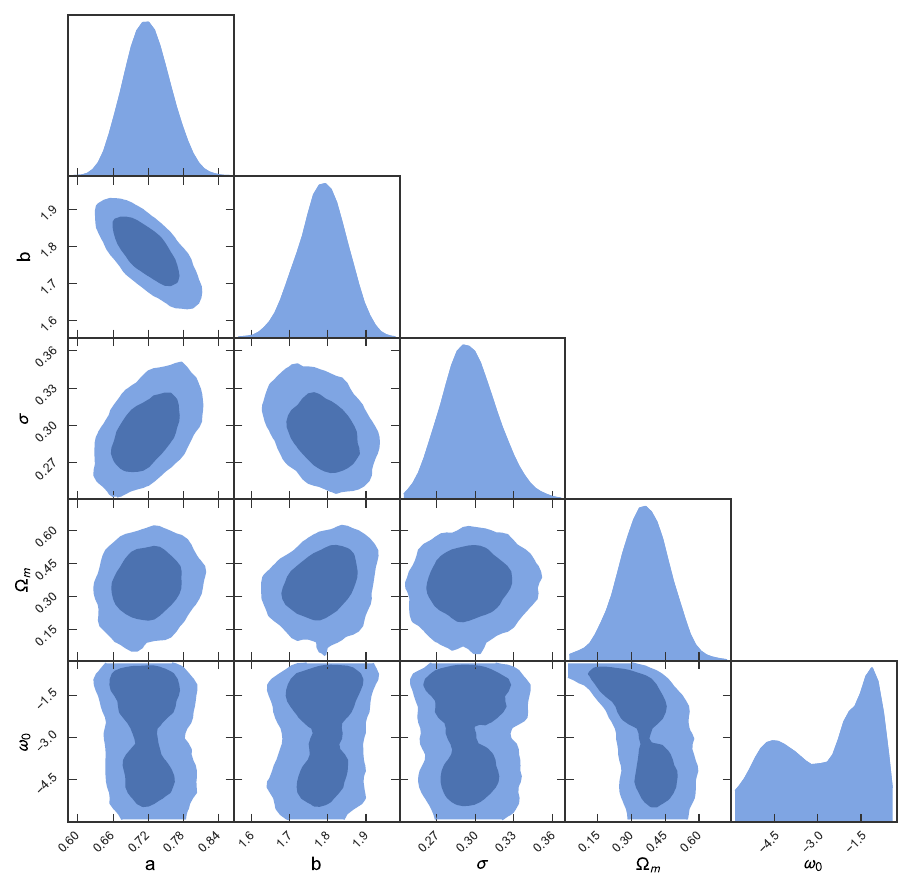}
\hfill
\includegraphics[width=0.48\hsize,clip]{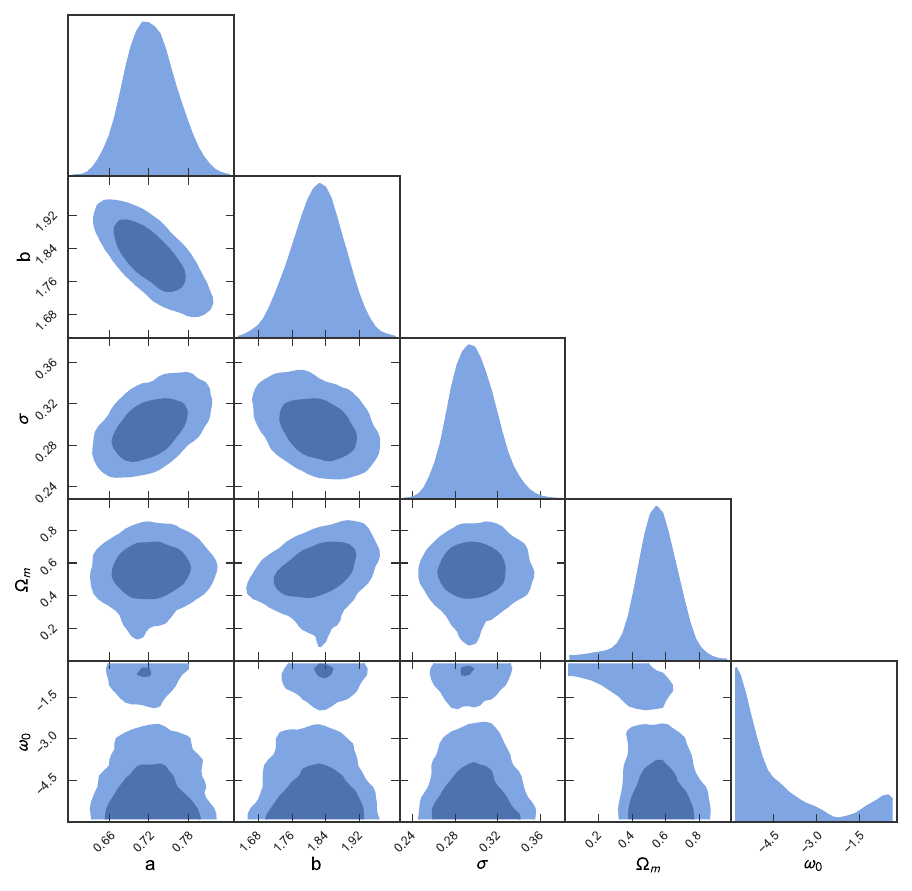}
\hfill}
\caption{Contour plots of the best-fit parameters for the $E_p-E_{iso}$ correlation and flat $\omega_0$CDM model without LRG1 (right) and with LRG1 (left).}
\label{fig:4}
\end{figure*}

\begin{figure*}
\centering
{\hfill
\includegraphics[width=0.48\hsize,clip]{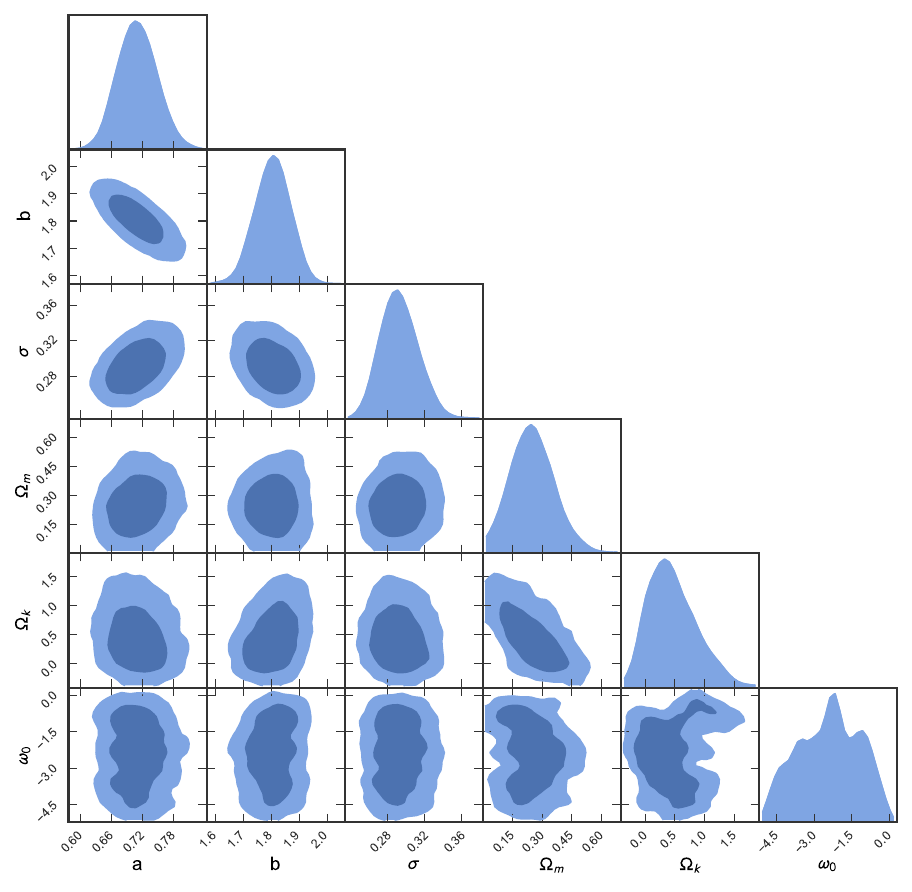}
\hfill
\includegraphics[width=0.48\hsize,clip]{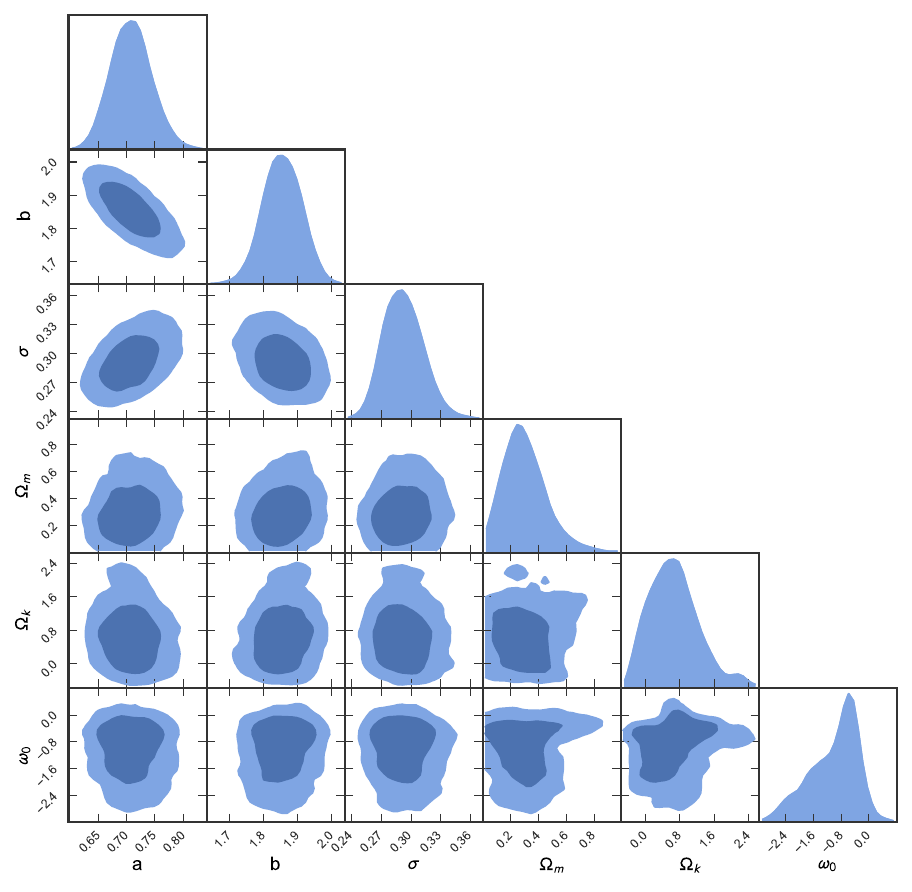}
\hfill}
\caption{Contour plots of the best-fit parameters for the $E_p-E_{iso}$ correlation and non-flat $\omega_0$CDM model without LRG1 (right) and with LRG1 (left).}
\label{fig:5}
\end{figure*}

\begin{figure*}
\centering
{\hfill
\includegraphics[width=0.48\hsize,clip]{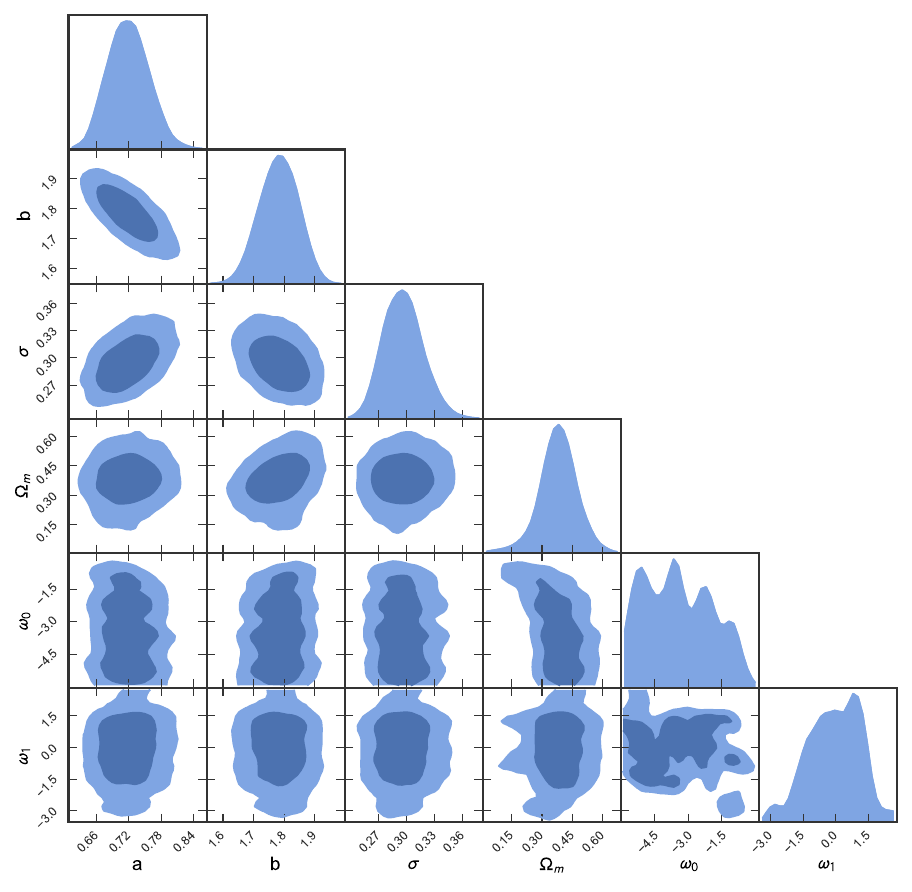}
\hfill
\includegraphics[width=0.48\hsize,clip]{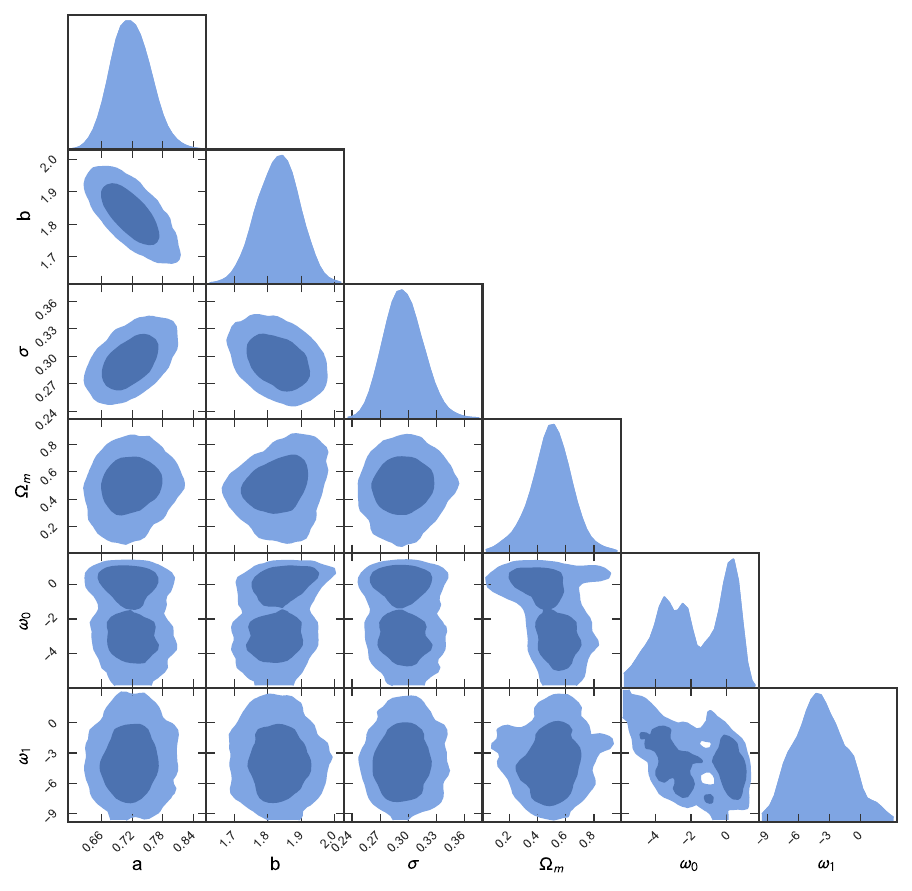}
\hfill}
\caption{Contour plots of the best-fit parameters for the $E_p-E_{iso}$ correlation and flat $\omega_0\omega_1$CDM model without LRG1 (right) and with LRG1 (left).}
\label{fig:6}
\end{figure*}

\begin{figure*}
\centering
{\hfill
\includegraphics[width=0.48\hsize,clip]{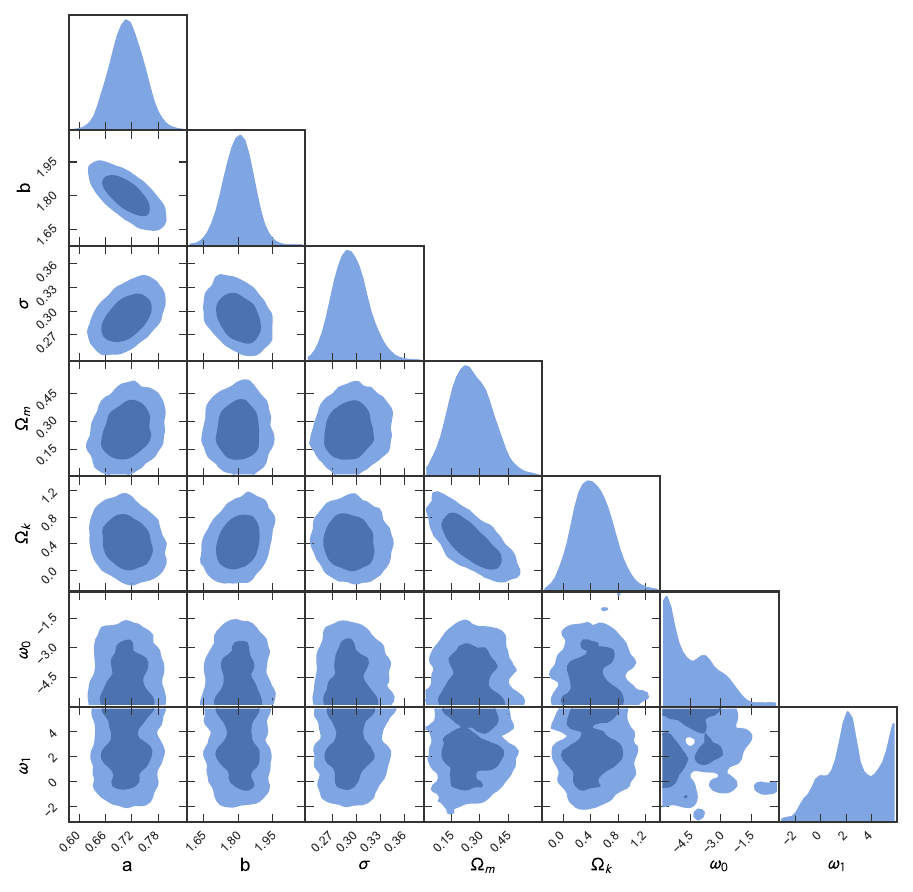}
\hfill
\includegraphics[width=0.48\hsize,clip]{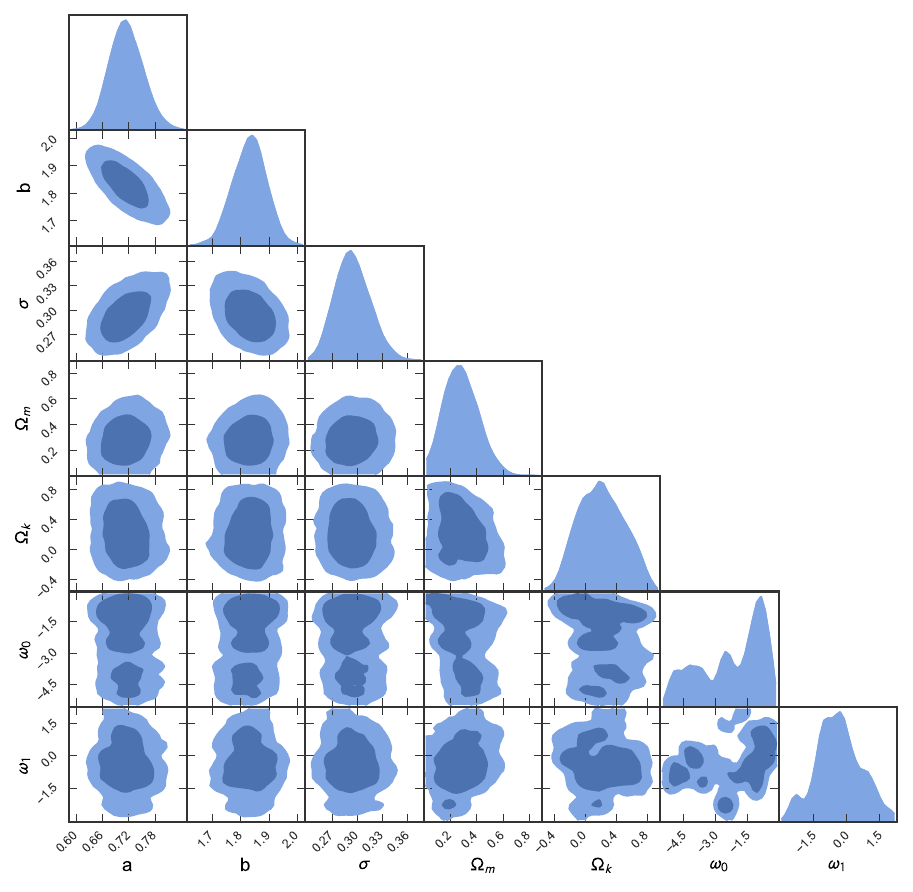}
\hfill}
\caption{Contour plots of the best-fit parameters for the $E_p-E_{iso}$ correlation and non-flat $\omega_0\omega_1$CDM model without LRG1 (right) and with LRG1 (left).}
\label{fig:6}
\end{figure*}

    \end{widetext}

\end{document}